\pdfoutput=1
\documentclass{aa} 
\usepackage{graphicx}
\usepackage[super]{nth}
\usepackage{txfonts}
\usepackage{natbib}
\usepackage{float}
\usepackage{amsmath}
\usepackage{esint}
\usepackage{subfig}
\newcommand\tab[1][1cm]{\hspace*{#1}}
\usepackage{multirow}
\usepackage[percent]{overpic}
\usepackage{xcolor}
\graphicspath{{./figures/}}
\bibpunct{(}{)}{;}{a}{}{,} 
\begin{document}

        \title{Numerical simulations of shear-induced consecutive coronal mass ejections}
        
        \author{D.-C. Talpeanu \inst{1,2}
                \and
                E. Chan\'{e}\inst{1}                    
                \and
                S. Poedts \inst{1,3}
                \and
                E. D'Huys\inst{2}
                \and
                M. Mierla \inst{2,4}
                \and
                I. Roussev \inst{5,1}
                \and
                S. Hosteaux \inst{1}
        }
        
        \institute{Centre for mathematical Plasma Astrophysics (CmPA), Department of Mathematics, KU Leuven, Celestijnenlaan 200B, 3001 Leuven, Belgium\\
                \email{dana.talpeanu@observatory.be}
                \and
                SIDC - Royal Observatory of Belgium (ROB), Av. Circulaire 3, 1180 Brussels, Belgium               
                \and
                Institute of Physics, University of Maria Curie-Sk{\l}odowska, PL-20-031 Lublin, Poland
                \and 
                Institute of Geodynamics of the Romanian Academy, Jean-Louis Calderon 19-21, 020032 Bucharest, Romania
                \and 
                Division of Atmospheric and Geospace Sciences - Directorate of Geosciences - National Science Foundation, Arlington, Virginia, USA        
        }
        
        \date{Received : to be filled in; accepted: to be filled in}
        
        
        \abstract
        {It is widely accepted that photospheric shearing motions play an important role in triggering the initiation of coronal mass ejections (CMEs). Even so, there are events for which the source signatures are difficult to locate, while the CMEs can be clearly observed in coronagraph data. These events are therefore called `stealth' CMEs. They are of particular interest to space weather forecasters, since eruptions are usually discarded from arrival predictions if they appear to be backsided, which means not presenting any clear low-coronal signatures on the visible solar disc. Such assumptions are not valid for stealth CMEs since they can originate from the front side of the Sun and be Earth-directed, but they remain undetected and can therefore trigger unpredicted geomagnetic storms.}
        {We numerically model and investigate the effects of shearing motion variations onto the resulting eruptions and we focus in particular on obtaining a stealth CME in the trailing current sheet of a previous ejection.}
        {We used the 2.5D magnetohydrodynamics
 (MHD) package of the code MPI-AMRVAC to numerically simulate consecutive CMEs by imposing shearing motions onto the inner boundary, which represents, in our case, the low corona. The initial magnetic configuration consists of a triple arcade structure embedded into a bimodal solar wind, and the sheared polarity inversion line is found in the southern loop system. The mesh was continuously adapted through a refinement method that applies to current carrying structures, allowing us to easily track the CMEs in high resolution, without resolving the grid in the entire domain. We also compared the obtained eruptions with the observed directions of propagation, determined using a forward modelling reconstruction technique based on a graduated cylindrical shell (GCS) geometry, of an initial multiple coronal mass ejection (MCME) event that occurred in September 2009. We further analysed the simulated ejections by tracking the centre of their flux ropes in latitude and their total speed. Radial Poynting flux computation was employed as well to follow the evolution of electromagnetic energy introduced into the system.}
        {Changes within 1\% in the shearing speed result in three different scenarios for the second CME, although the preceding eruption seems insusceptible to such small variations. Depending on the applied shearing speed, we thus obtain a failed eruption, a stealth, or a CME driven by the imposed shear, as the second ejection. The dynamics of all eruptions are compared with the observed directions of propagation of an MCME event and a good correlation is achieved. The Poynting flux analysis reveals the temporal variation of the important steps of eruptions.}
        {For the first time, a stealth CME is simulated in the aftermath of a first eruption, originating from an asymmetric streamer configuration, through changes in the applied shearing speed, indicating it is not necessary for a closed streamer to exist high in the corona for such an event to occur. We also emphasise the high sensitivity of the corona to small changes in motions at the photosphere, or in our simulations, at the low corona.}
        
        \keywords{magnetohydrodynamics (MHD) --
                methods: numerical --
                Sun: coronal mass ejections (CMEs) --
                methods: observational
        }
        
\maketitle


        \section{Introduction} \label{sec:intro}

Coronal mass ejections (CMEs) are some of the most energetic solar events that expel magnetic field-loaded plasma into the interplanetary space. These events release energies of the order of 10$^{32}$ ergs \citep[e.g.][]{forbes_energies} and they leave the solar environment with speeds between 20 km s$^{-1}$
 and 2500 km s$^{-1}$ \citep{webb_speeds}. If headed towards Earth, CMEs can provoke serious geomagnetic disturbances, induce electrical currents in power grids, disrupt the global positioning system, and endanger the life of astronauts. One way of mitigating these effects is to forecast the arrival of the associated interplanetary coronal mass ejection (ICME) at our planet, by identifying its source, direction, and speed. If source signatures such as flares, filament eruptions, coronal dimmings, or extreme ultraviolet (EUV)  waves are not clearly visible on the side of the Sun directed towards Earth, it is usually assumed that the event is backsided. However, there have been reported cases of eruptions lacking clear source locations, which were still Earthward directed. These situations are extremely puzzling for space weather forecasters, since geomagnetic storms caused by such CMEs can occur unpredicted and can potentially cause disruptions. Ejections without clear low coronal signatures are referred to as `stealth' CMEs and were first investigated by \citet{robbrecht_stealth}. The authors reported a slow CME eruption caused by a streamer blowout on 1-2 June 2008, which spanned only 54$\degr$ in angular width and lacked the typical low-coronal signatures. Nevertheless, using the multi-viewpoint capability provided by the twin Solar Terrestrial Relations Observatory (STEREO) spacecraft \citep{stereo_mission} and the onboard coronagraphs COR1 and COR2 of the Sun Earth Connection Coronal and Heliospheric Investigation (SECCHI) imaging suite \citep{stereo_secchi}, these authors reconstructed the source of the CME and located it in a quiet-Sun front-sided region, contrary to the intuitive assumptions. The authors attributed the absence of traces to the large lift-off height and concluded that a CME originating from above 1.4 solar radii (R$_{\sun}$) should not leave any associated dimming, hence creating a stealth eruption. Other studies reporting CMEs without low coronal signatures include those of \citet{ma_stealth}, \citet{elke_stealth}, \citet{kilpua_stealth}, and \citet{nariaki_stealth}. Most of the studies on such events agree on two of their prevalent properties: a small angular width, of mostly $<60\degr$, and low initiation and propagation speeds, which are usually below 300 km s$^{-1}$. Another study conducted by \citet{alzate_sources} used advanced image processing techniques and located sources for the eruptions presented in \citet{elke_stealth}, confirming the low kinetic energies observed so far. These results emphasise the issues faced by space weather forecasters, which are the need for better imaging instruments, continuous observations from different vantage points, and a better understanding of the causes of such problematic eruptions. \\          
Whether CMEs have a clear source region is closely related to the eruption mechanism, the height of reconnection, the overlying magnetic field, the limitation of current instrument capabilities \citep{howard_harrison_review}, and other solar atmospheric conditions. It is generally accepted within the solar community that photospheric shearing motions present one of the driving mechanisms for CMEs, as they increase the magnetic helicity and consequently, the free magnetic energy. \citet{linker_shear}, \citet{bart_shear, bart_breakout}, \citet{devore_shear}, and \citet{lynch_stealth} are just some examples of studies using numerical simulations with imposed shearing motions to obtain CMEs and investigate their dynamical properties, ranging from the simplest 2.5D bipolar magnetic configurations, to full Sun 3D multi-polar structures. \\
Proposed mechanisms for stealth CME initiation include streamer blowout \citep[e.g.][]{robbrecht_stealth}, empty filament channel eruption \citep[e.g.][]{pevtsov_filament}, and coronal magnetic field reconfiguration, usually in the trailing current sheet of a first CME \citep[e.g.][]{bemporad}. We modelled and analysed the latter type of event by means of shearing motions imposed at the lower boundary (low corona), and also focussed on the effect of variation of the shear amplitude on the resulting eruptions.\\     
It could be  argued that this particular type of stealth events fits in the `plasma blobs' category, as they originate in the current sheet created behind a first CME. These blobs were observed and analysed by several authors, including \citet{webb_and_cliver_blobs}, \citet {ko_blobs}, and \citet{webb_and_vourlidas_blobs}, who reported such narrow structures flowing along the current sheets that form after fast CMEs; these blobs were also numerically simulated by \citet{riley_sim_blobs}. Whether these blobs can be called CMEs or not is still a topic of debate, but their eruption mechanism is very similar, and therefore, for this study the distinction is not important.\\       
This work is similar to that of \citet{zuccarello} and \citet{bemporad} owing to the initial magnetic configuration and shearing profile imposed, but the main difference is that we did not require a change in magnetic field strength to obtain the second eruption. Instead, small changes applied to the shearing speed resulted in the coronal magnetic field reconfiguration, creating the second stealthy CME. Furthermore, as opposed to having a fixed grid, we use an adaptive mesh refinement (AMR) routine that increases the resolution only in regions of interest, which are areas where the electrical current is enhanced. 
        

        \section{Observations} \label{sec:observations} 
        
\begin{figure}[h!]
                \centering
                \includegraphics[width=0.8\columnwidth]{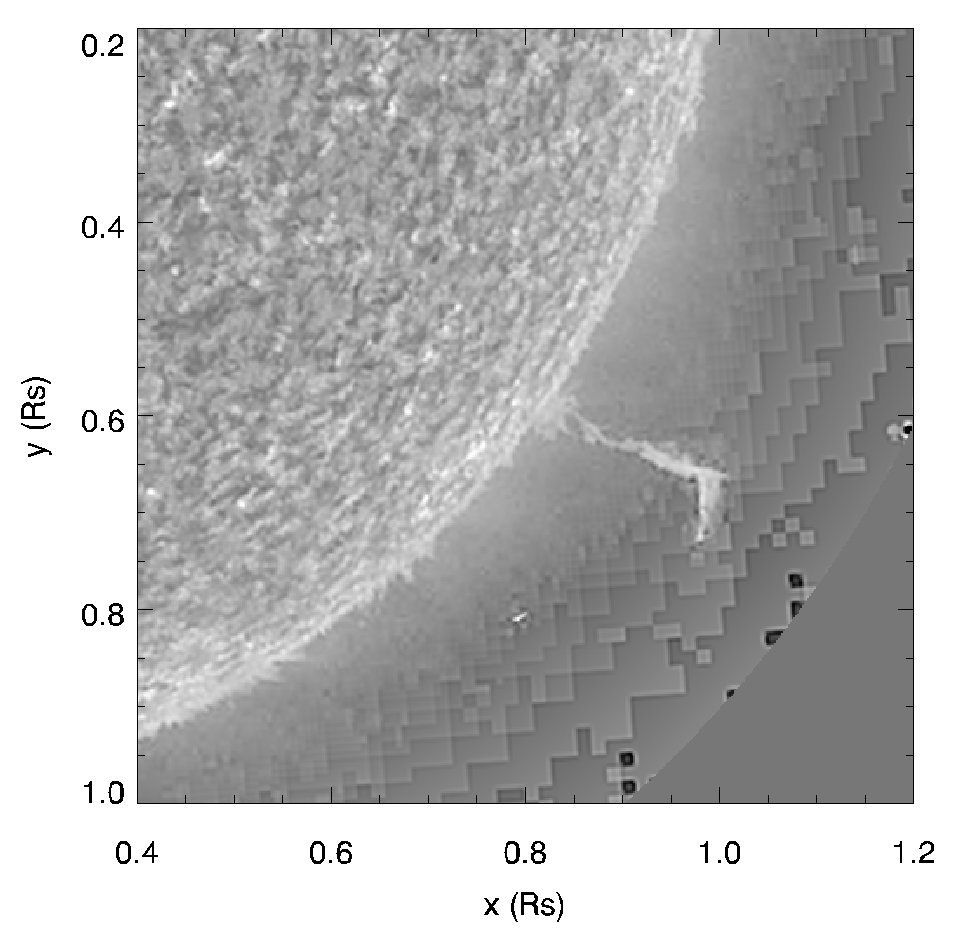}
                \caption{Image from STEREO-B EUVI (304) of the erupting prominence, taken on 21 September 2009, at 16:36UT. The image was scaled to enhance the prominence and a mask was applied from 1.35 R$_{\sun}$ outwards to remove the large noise at the edges.}
                \label{fig:prominence}
        \end{figure}

        
        The observational basis of this work consists of a multiple coronal mass ejections event \citep[MCMEs;][]{bemporad} that occurred on 21-22 September 2009. The first CME (hereafter CME1) was associated with a western limb prominence eruption and originated from an approximate latitude of 37$\degr$ south, as seen by SECCHI-Extreme UV Imager \citep[EUVI;][]{secchi} on board the STEREO-B spacecraft. This CME was correlated with a small active region located at 38\degr S and 15\degr W, which is not recorded in the NOAA catalogue. The ejected material left the EUVI field of view on  21 September, at around 19:37 UT, as seen in Fig. \ref{fig:prominence}. The second CME (hereafter CME2) entered the COR1-B \citep{thompson-cor, stereo_secchi} coronagraph field of view on 22 September 2009 at $\sim$04:05 UT, almost eight hours after CME1. We investigated running difference images in different wavelengths and with different time steps, but we could not identify any plasma motions that could be distinguished clearly from the background noise and that would indicate the origin of the second CME. Because of the difficulty of locating the source of CME2, the lack of obvious low coronal signatures, and the fact that it was visible only when reaching higher distances from the solar surface, we consider it a stealth CME. Even though it was fainter and narrower than CME1, CME2 propagated very similarly to the first event, originating from approximately the same coronal region and presenting a strong equatorial deflection. This behaviour was explained and numerically simulated by \cite{zuccarello}. Their paper also contains a more detailed description of the eruptions and positions of STEREO-A and B spacecraft during the event.      
        
\begin{figure*}[h!]    
                \centering
                First eruption \tab[7cm] Second eruption\\      
                \begin{tabular}{c c}
                        \hspace*{-0.17in}
                        \parbox[t]{0.5\linewidth}
                        {
                                \centering
                                \label{fig:cme1_cor2a}{\includegraphics[width=0.48\linewidth]{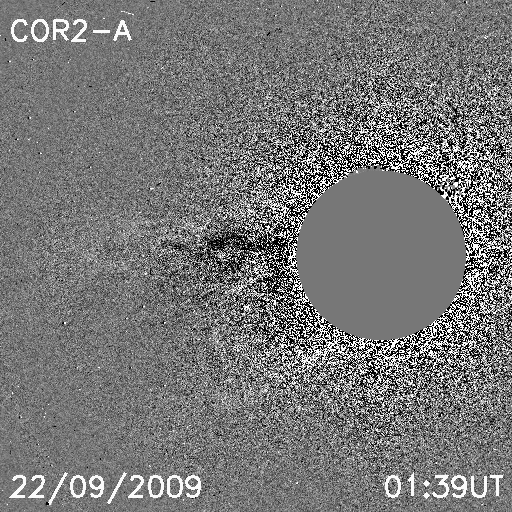}}
                                \label{fig:cme1_cor2b}{\includegraphics[width=0.48\linewidth]{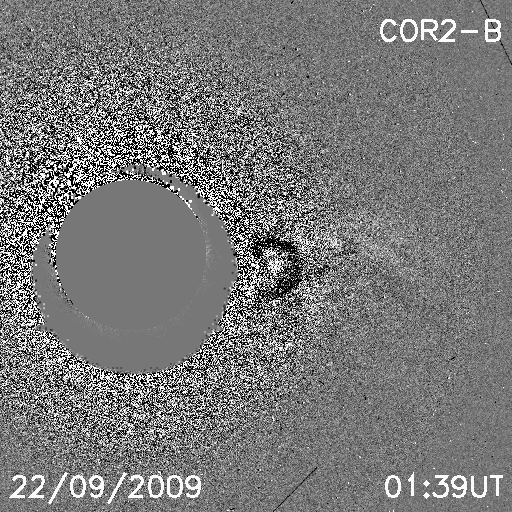}}
                                \label{fig:cme1_gcs_a}{\includegraphics[width=0.48\linewidth]{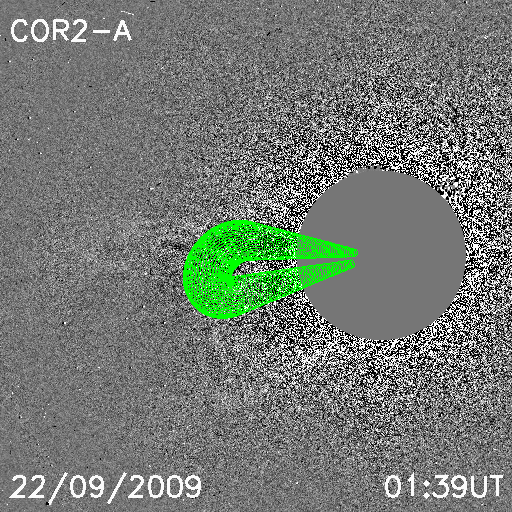}}
                                \label{fig:cme1_gcs_b}{\includegraphics[width=0.48\linewidth]{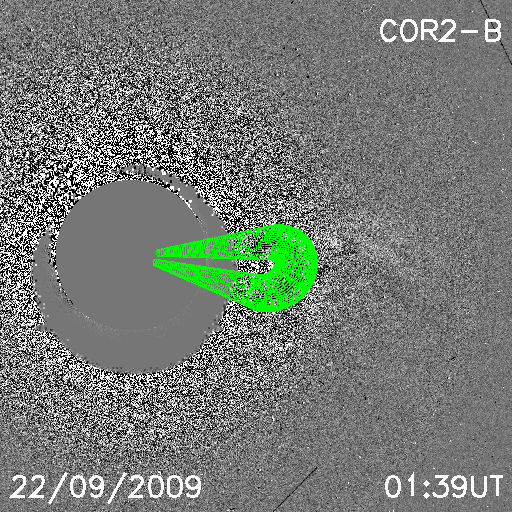}}                   
                        }
                        \hspace*{-0.04in}                       
                        \vline
                &
                        \hspace*{-0.18in}
                        \parbox[t]{0.5\linewidth}
                        {       
                                \centering      
                                \label{fig:cme2_cor2a}{\includegraphics[width=0.48\linewidth]{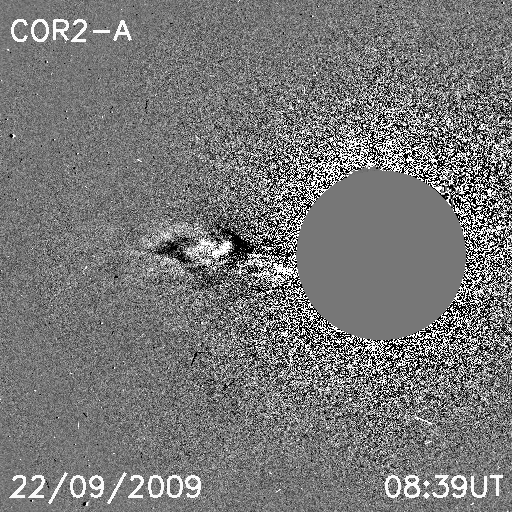}}
                                \label{fig:cme2_cor2b}{\includegraphics[width=0.48\linewidth]{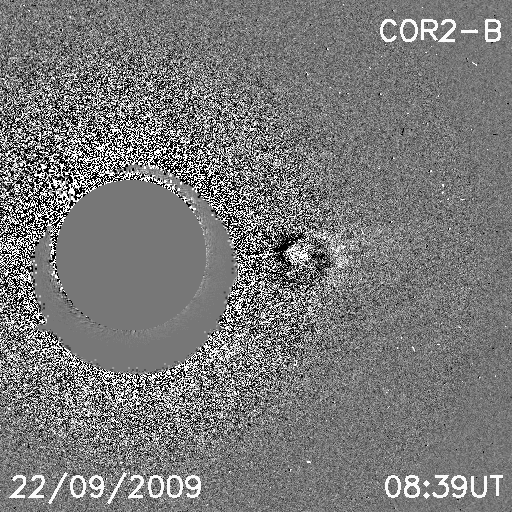}}
                                \label{fig:cme2_gcs_a}{\includegraphics[width=0.48\linewidth]{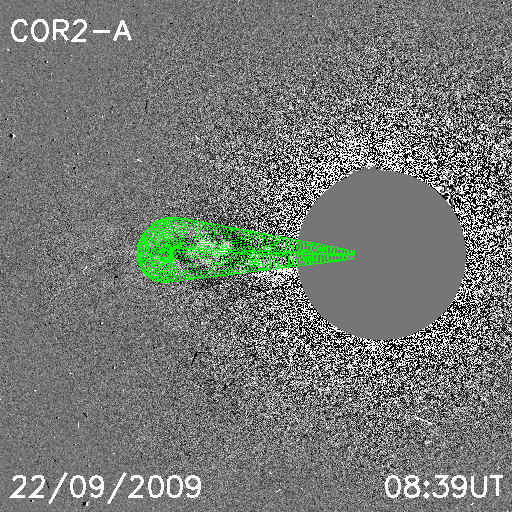}}
                                \label{fig:cme2_gcs_b}{\includegraphics[width=0.48\linewidth]{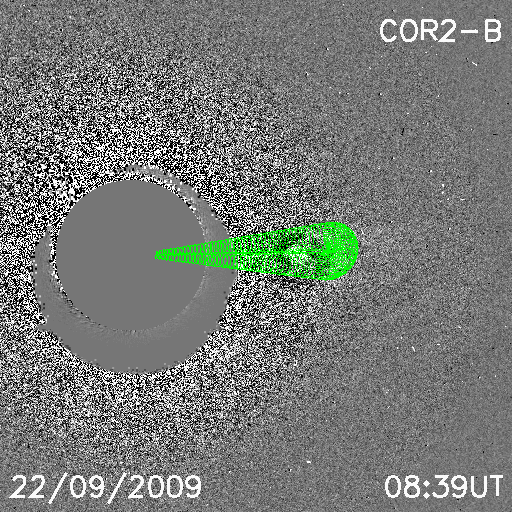}}                   
                        }                       
                \end{tabular}
                \caption{Top: Running difference images from COR2 used in the GCS reconstruction model. Bottom: Reconstructions (green wire) using GCS superposed onto the upper corresponding panels in the case of CME1 (left) and CME2 (right).}
                \label{fig:gcs_reconstructions}
        \end{figure*}

        \citet{bemporad} calculated the velocities of both CMEs within 4 R$_{\sun}$ using height versus time maps (`ht-plots'), and obtained values below 164 km s$^{-1}$. Since we were more interested in the direction of propagation of the two CMEs after the deflection, we performed 3D reconstructions using the forward modelling technique higher in altitude (above 6 R$_{\sun}$), in the COR2 field of view. The geometrical shape chosen for the CME fitting was the graduated cylindrical shell (GCS) model \citep{gcs_thernisien_2006, gcs_thernisien_2009, gcs_thernisien_2011} available as a SolarSoftWare (SSW) package\footnote{http://www.lmsal.com/solarsoft/}, which approximates the CME structure to a bent cylinder connected to the Sun by two cones, representing the CME front and its footpoints, respectively. The shape can be changed through six free parameters, which we adjusted such that the predefined configuration matches the eruption in white-light running difference images (Fig. \ref{fig:gcs_reconstructions}). For a more accurate computation, the reconstruction was performed when the CMEs were best seen in the observations, i.e. at two different times for CME1, and at three times for CME2. Table \ref{table:gcs_results} shows the resulting longitude, latitude, and height in the Stonyhurst system of coordinates \citep{thompson_stonyhurst}. All five measurements were made for the same day, that is 22 September 2009. 
        
           \begin{table}[h!]
                \centering
                \caption[]{Results from the GCS analysis.}
                \begin{tabular}{c c c c c} 
                \label{table:gcs_results}
                Eruption              & Hour     &  Long.     & Lat.      & Height/{[R$_{\sun}$]} \\
                \hline 
                \noalign{\smallskip}
        \multirow{2}{*}{CME1} & 01:39 UT &  6.1\degr & 2.2\degr & 6.9    \\
                                              & 02:39 UT &  5.6\degr & 2.8\degr  & 7.6    \\
                \noalign{\smallskip}                          
            \hline
            \noalign{\smallskip}
                \multirow{3}{*}{CME2} & 07:54 UT &  6.7\degr  & 7.8\degr  & 6.9    \\
                                                          & 08:39 UT &  5.6\degr & 7.8\degr  & 8.4    \\
                                                          & 12:24 UT &  7.8\degr & 5.6\degr & 14.5  \\
                \noalign{\smallskip}
            \hline 
            \end{tabular}
           \end{table}

\begin{figure}[h!]
                \centering
                \includegraphics[width=1\columnwidth]{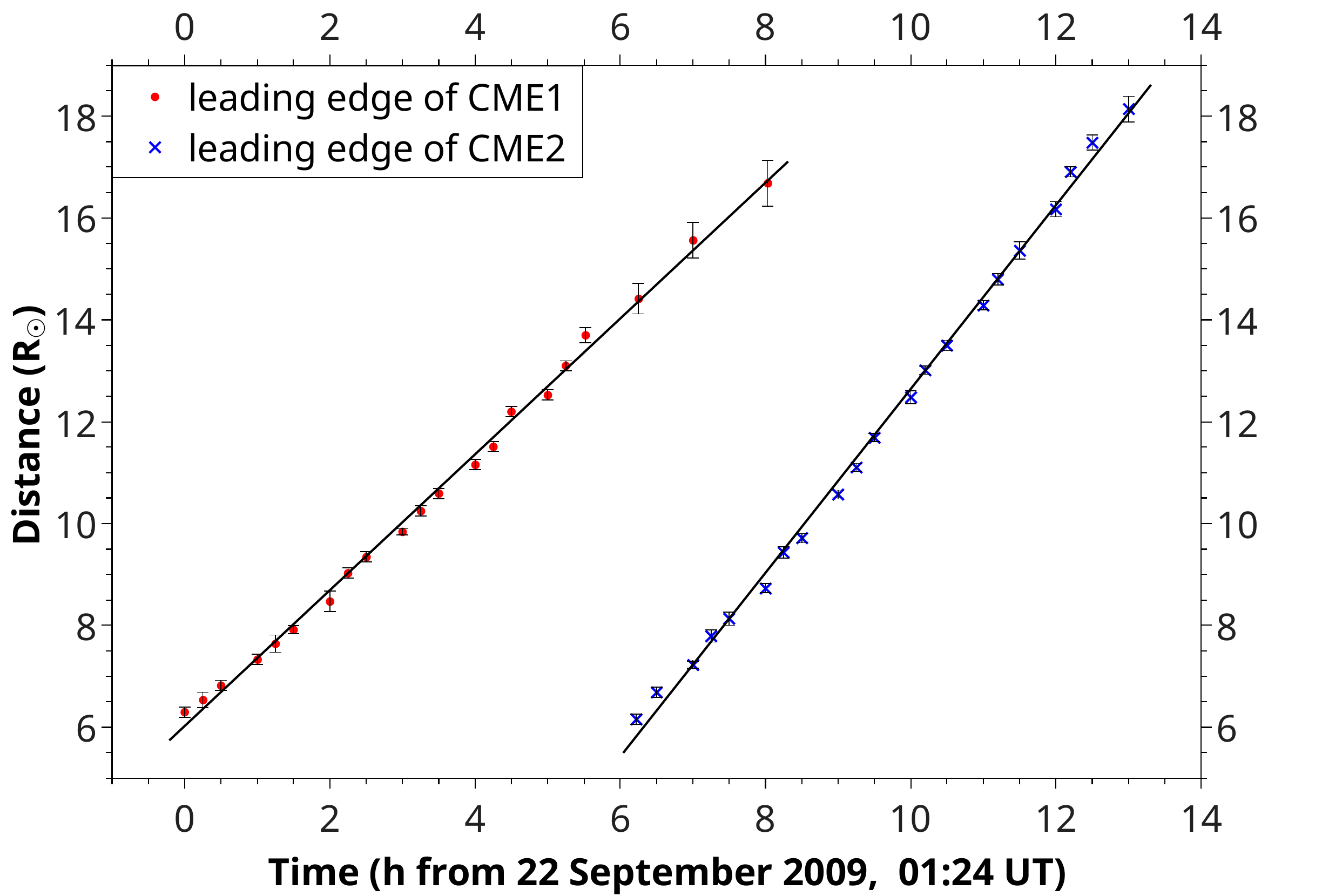}
                \caption{Height-time plots of the two studied CMEs, erupting between 21-22 September 2009, based on STEREO-B coronagraph observations.}
                \label{fig:h-t plots}
                \vspace{-0.3cm}
        \end{figure}

\begin{figure}[h!]
                \centering
                \includegraphics[width=1\columnwidth]{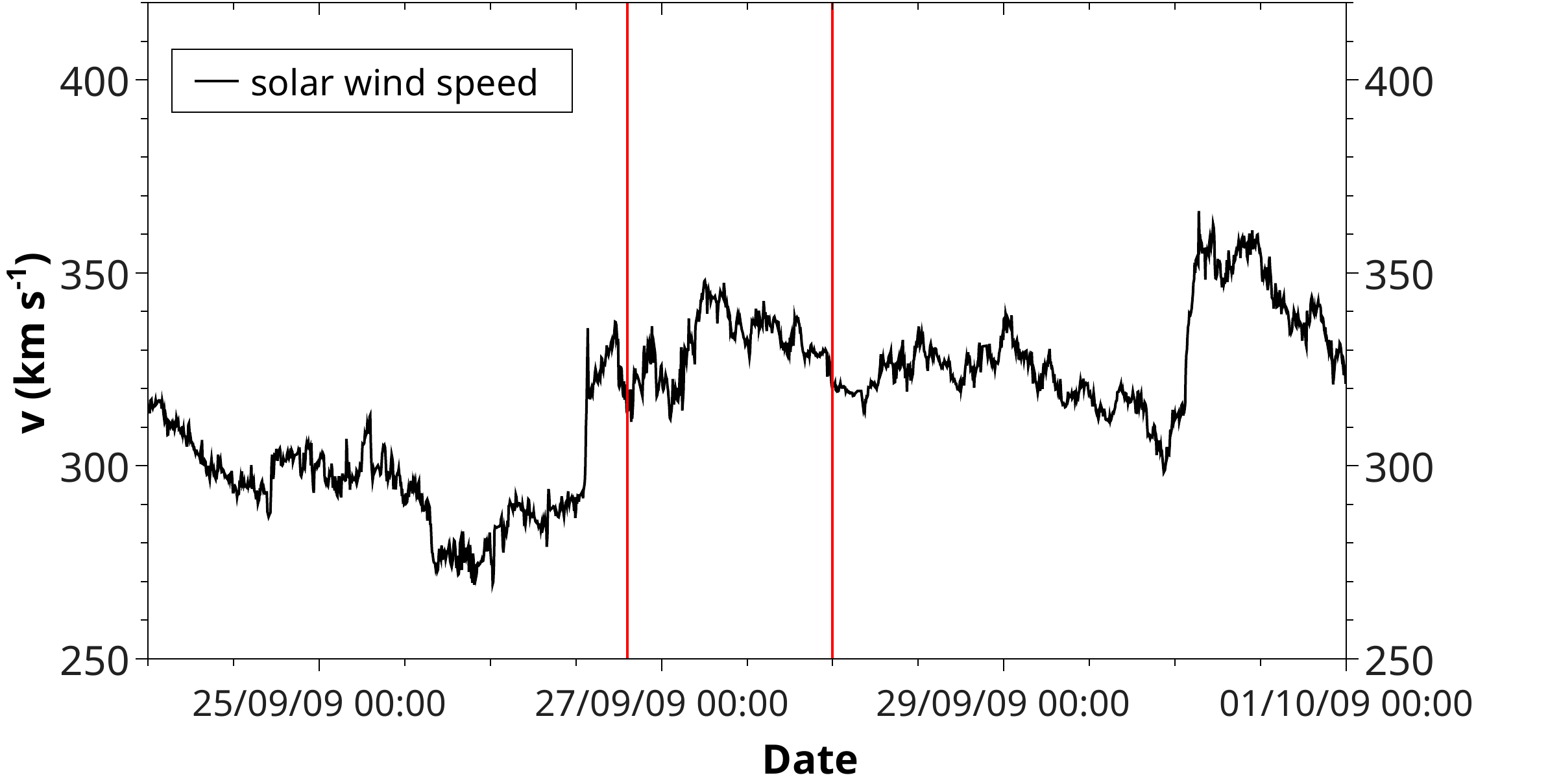}
                \caption{Solar wind speed recorded at Earth, taken from the OMNI database. The two red lines indicate the expected arrival time interval of the two CMEs.}
                \label{fig:speed_5min_omni}
                \vspace{-0.3cm}
        \end{figure}

        The deprojected speed of a CME is mainly influenced by its longitude, which in this case had average values of 5.82\degr W for CME1, and 6.7\degr W for CME2. The smallest longitude value found for CME2 (5.6\degr, see Table \ref{table:gcs_results}) is probably due to errors from the reconstruction method and not a real variation. We previously measured the projected heights of the fronts of both CMEs in COR2-B plane of sky, when the angle between STEREO-B and the Sun-Earth line was $-55.6\degr$. We thus computed the deprojected distances of both CME fronts from the Sun centre. We plotted the deprojected height of the CME1 leading edge as a function of time from 6.3 R$_{\sun}$ to 16.7 R$_{\sun}$ (see Fig. \ref{fig:h-t plots}) and fitted the data with a first order polynomial, resulting in a deprojected velocity of 257 $\pm$ 69  km s$^{-1}$. This speed is higher than 164 km s$^{-1}$ measured in the low corona within 4 R$_{\sun}$, and this acceleration above 6 R$_{\sun}$ is also seen in the automated CACTus CME list \citep{cactus} based on COR2-B observations, where the velocity is listed as 277 km s$^{-1}$. As seen in the ht-plot, CME2 was faster and propagated in the same field of view with a deprojected velocity of 349 $\pm$ 70 km s$^{-1}$, mainly because it was expelled into a depleted solar wind following the passage of CME1. Given these slow speeds, a good assumption would be that the CMEs eventually reached the slow solar wind speed (as seen in Fig. \ref{fig:speed_5min_omni}) due to the drag force exerted on them. The error bars in Fig. \ref{fig:h-t plots} were calculated by selecting the CME front several times and extracting the largest variation between measurements. We computed the standard deviation in speed using the derivsig.pro function available in IDL, based on the height and time measurements and their error bars. The function computes the speed uncertainty values from the errors in height using an error propagation formula.      

        In order to obtain the speed of the wind into which the CMEs were ejected, we used the in-situ measured solar wind speed as extracted from NASA/GSFC's OMNI data set through OMNIWeb\footnote{https://omniweb.gsfc.nasa.gov/} between 3 to 8 days after the eruption. This parameter is needed as input for the simulations described in Section \ref{sec:simulations}. In this time interval, we carried out a back-calculation for each OMNI speed measurement the date and time at which the embedded CMEs should have originated from the Sun, in the assumption they had approximately the same velocity as the solar wind. We then compared this with the real eruption time and concluded the arrival time of the CMEs should be 27-28 September, as indicated by the red vertical lines in Fig. \ref{fig:speed_5min_omni}. In this period, the solar wind speed did not present any major jumps, as expected owing to the low speeds of the CMEs, and had an average value of $\approx$330 km s$^{-1}$. The small jump in speed on 26 September at 13:00 is not correlated to any other in situ ICME signature and this jump also arrived too early, therefore it was not considered to be the beginning of the ICME. We performed a more detailed study of other solar wind  parameters as well, such as magnetic field components, to determine the arrival of discussed CMEs at Earth, and this topic shall be the focus of a follow-up paper.

        \section{Simulations set-up and method} \label{sec:simulations}
        
        Our goal was to investigate the effect of variations in the amplitude of shearing motions on the dynamics of CMEs, starting from the observed event discussed in Section \ref{sec:observations}. For this purpose, we performed 2.5D spherical axisymmetric numerical simulations using the Message Passing Interface - Adaptive Mesh Refinement - Versatile Advection Code \citep[MPI-AMRVAC;][]{keppens_amrvac, porth_amrvac, xia_amrvac}, which resolves the system of partial differential equations for magnetohydrodynamics (MHD) on a 2D logarithmically stretched grid. The mesh has an initial resolution of 480 $\times$ 240 cells and the domain spans between the north and south poles of the Sun, and the Earth, that is (\emph{r}, $\theta$) $\in$ [1, 215] R$_{\sun}$ $\times$ [0\degr, 180\degr], where $r$ is the radial distance from the centre of the Sun and $\theta$ is the heliographic colatitude. Even though the grid extends until 1 AU, in this study we only focussed on the near-Sun region, below 20 R$_{\sun}$. The propagation of the simulated CMEs to Earth will be discussed in a follow-up paper. In order to achieve much better detail of the dynamics of the eruption and of the structures that form, without increasing the resolution of the entire domain, an AMR protocol was used. The grid was refined to a maximum of two additional levels and only in the regions of interest, where the electrical current presented an enhancement, that is at potential magnetic reconnection sites. This increase is quantified through a parameter taken from \citet{karpen} and also used by \citet{skralan} as follows:
        
        \begin{equation}
        c \equiv \frac{|\iint_S \nabla \times \boldsymbol{B} \cdot \mathrm{d} \boldsymbol{a}|}{\oint_C |\boldsymbol{B} \cdot \mathrm{d} \boldsymbol{l}|} = \frac{|\oint_C \boldsymbol{B} \cdot \mathrm{d} \boldsymbol{l}|}{\oint_C |\boldsymbol{B} \cdot \mathrm{d} \boldsymbol{l}|} = \frac{|\sum_{n=1}^{4} B_{t,n} l_n|}{\sum_{n=1}^{4} |B_{t,n} l_n|}, \label{c_param}
        \end{equation}
        
        \noindent where $B_{t,n}$ represents the tangential component of the magnetic field along the segment $l_n$. The line integral at the numerator is obtained by applying the Stokes' theorem on the surface integral in the left hand-side term, and in this case the curve (contour) and the surface are those of a grid cell. In their discrete form, the line integrals are rewritten as the sum of the product between the tangential magnetic field and the length of the edges, along all four sides of a cell. Therefore, $c$ is simply the ratio of the magnitude of the electrical current passing through the surface $S$ spanning loop $C$ to the sum of the absolute value of all of its components. In this definition, the parameter $c$ varies from 0 to 1, depending on the non-potentiality of the magnetic field. The blocks of the grid are refined if and where $c$ crosses a certain threshold, which we chose as $c$ > 0.02. On the other hand, if $c$ < 0.01, the grid is coarsened to a lower level, since there are no strong current-carrying structures in that region. In between these two values, the AMR routine does not impose any constraints, so the grid retains its resolution from previous time steps. The refinement is also fixed to a maximum number of two levels (on top of the base grid) for distances below 2.5 R$_{\sun}$ and for $\theta$ $\in$ [27\degr, 153\degr], such that there is no change in diffusivity close to the boundary, which would have affected the eruption dynamics and introduced artificial phenomena. Because of the stretching of the grid, the scale of cells is kept constant, and the size ratio of the widths (or heights) of the farthest to closest cells to the inner boundary is $\approx$ 215 for the same mesh refinement level.\\        
        The MHD equations are temporally discretised using a two-step predictor corrector, which is suitable for a finite-volume discretisation method called the total variation diminishing Lax-Friedrichs (TVDLF) scheme. We used the most diffuse and stable type of slope limiter, minmod, and a CFL number of 0.3. The magnetic field solution is kept divergence free using the generalised Lagrange multiplier \citep[GLM; ][]{glm} method, which introduces a new variable into the system, that damps and transports the unphysical magnetic monopoles outside the computational domain. \\            
        The initial conditions (at the start of shear) consist of a realistic bimodal solar wind created by introducing extra source terms to the momentum and energy equations to account for gravity and heating mechanisms. This solar wind model is similar to that used in previous studies by \citet{jacobs_2005}, \citet{chane_2006}, \citet{chane_2008} and \citet{skralan_2019}. The volumetric heating function was initially used by \citet{groth} and \citet{manchester}, and takes the following empirical form:     
        
        \begin{equation}
        Q = \rho q_0 (T_0 - T) \exp \left[ -\frac{(r-1\mathrm{R_{\sun}})^2}{\sigma_0^2} \right],  
        \end{equation}
        
        \noindent where $\rho$ is the mass density, the amplitude of the volumetric heating $q_0=10^{6}$ ergs g$^{-1}$ s$^{-1}$ \rm{K}$^{-1}$, $r$ (R$_{\sun}$) is the distance from the centre of the Sun, and $T\,\mathrm{(K)}$ is the temperature. The values $T_0 \,\mathrm{(K)}$ and $\sigma_0$ (R$_{\sun}$) represent the target temperature and the heating scale height, respectively, and these are both dependent on the value of a critical angle, $\theta_0$ (measured from the North pole), as follows: from $\theta_0$ equator-ward, $T_0=1.32 \times 10^{6}\,\mathrm{K}$ and $\sigma_0=4.5\,\mathrm{R_\sun}$, and from $\theta_0$ pole-ward, $T_0=2.3 \times 10^{6}\,\mathrm{K}$ and $\sigma_0=4.5[2-\sin^2(\theta)/\sin^2(\theta_0)] \,\mathrm{R_\sun}$. To better reproduce the slow/fast solar wind features, $\theta_0$ is also defined with a distance dependence as follows:
        \begin{itemize}
                \item[$\bullet$] $\sin^2(\theta_0)=\sin^2(17.5 \degr)+\cos^2(17.5 \degr) (r-1\mathrm{R_{\sun}})/8\mathrm{R_{\sun}}$, for $r$ < 7 R$_{\sun}$,
                \item[$\bullet$] $\sin^2(\theta_0)=\sin^2(61.5 \degr)+\cos^2(61.5 \degr) (r-7\mathrm{R_{\sun}})/40\mathrm{R_{\sun}}$, for 7 R$_{\sun}$ $\leq$ $r$ < 47 R$_{\sun}$,  
                \item[$\bullet$] $\sin^2(\theta_0)= 1$, for 47 R$_{\sun}$ $\leq$ $r$.
        \end{itemize}
        
\begin{figure}[h!]
                        \centering
                        \begin{overpic}[width=0.9\columnwidth]{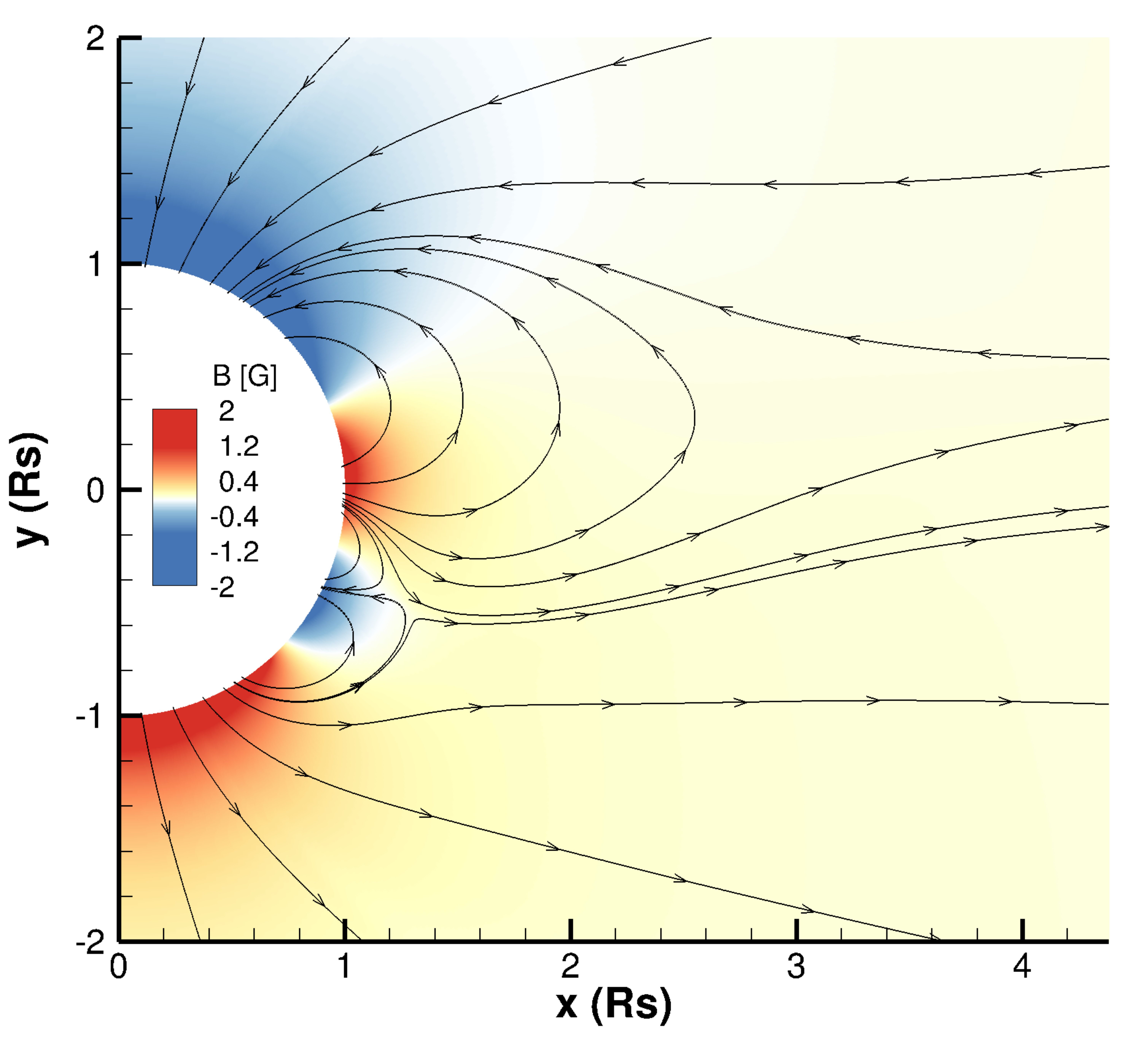}
                \put (14,58){\colorbox{white}{\parbox{0.07\linewidth}{\tiny $\mathrm{B_r [G]}$}}}
            \end{overpic}
                        \caption{Initial magnetic field configuration - $B_r$ (colour scale) and selected magnetic field lines in the meridional plane.}
                        \label{fig:arcades}
                \end{figure}

\begin{figure}[h!]
                        \centering
                        \includegraphics[width=0.9\columnwidth]{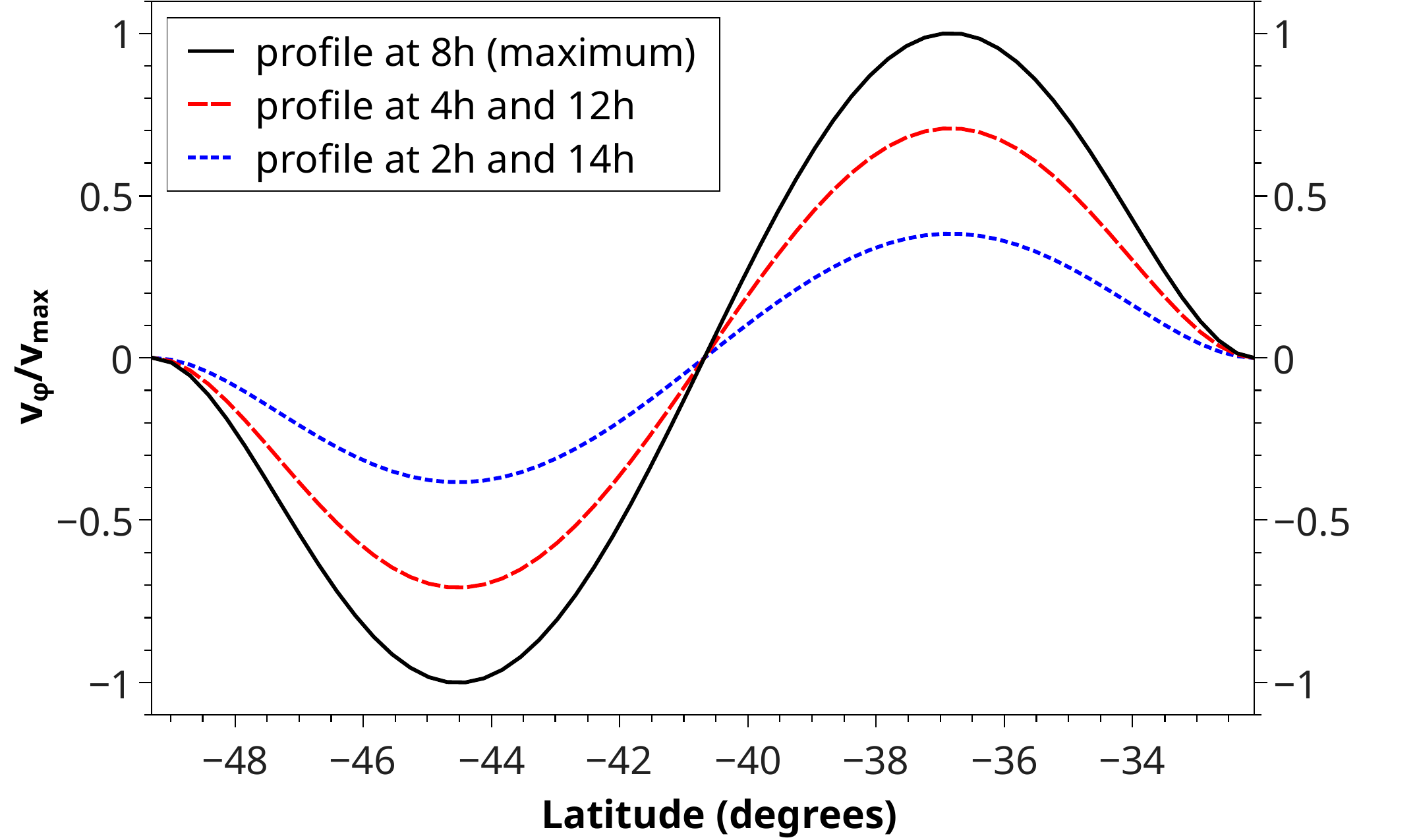}
                        \caption{Normalised shearing profile ($v_\phi/v_{max}$) as a function of latitude, at different simulation times from the start of shear.}
                        \label{fig:shear}
                \end{figure}

        The temperature and mass density are fixed at the inner boundary to $1.32 \times 10^{6}\,\mathrm{K}$ and $1.66 \times 10^{-16}\,\mathrm{g \, cm^{-3}}$, respectively. The temperature value at the boundary is chosen based on the impact it has on the solar wind speed through the momentum equation, provided that the observed speed at Earth was $\approx$330 km s$^{-1}$ (as discussed in Section \ref{sec:observations}). Therefore, the temperature is adjusted to provide a realistic slow solar wind speed at Earth of $\approx$330 km s$^{-1}$, and a fast solar wind speed at the poles of $\approx$735 km s$^{-1}$. The radial component of the momentum is extrapolated in the ghost cells, while the latitudinal component ($v_\theta$) is set to 0 at the boundary. The azimuthal component ($v_\phi$) is set such that it resembles the differential rotation of the Sun. In order to introduce a dipole field, the term $r^2B_r$ is fixed at the inner boundary, while $r^5B_\theta$ and $B_\phi$ are extrapolated from the first inner cell. At the outer supersonic boundary, the variables $r^2\rho$, $r^2\rho v_r$, $\rho v_\theta$, $r v_\phi$, $r^2B_r$, $B_\theta$, $r B_\phi$, and $T$ are continuous as well. \\
        A commonly encountered magnetic field structure on the Sun depicts a triple arcade system embedded in a helmet streamer (see Fig. \ref{fig:arcades}), which was also simulated by \citet{bemporad} and \citet{zuccarello} to study the deflection and dynamics of the chosen multiple event, and by \citet{karpen} to analyse a breakout event. We applied the same configuration in our model, by taking the curl of the following vector potential:
        
        \begin{equation}
        A_\phi = \frac{A_0}{r^4 \sin\theta} \cos^2 \left[ \frac{180\degr (\lambda+11.5\degr)}{2\Delta a}\right],  
        \end{equation}
        
        \noindent through which we obtained the magnetic field components
        
        \begin{eqnarray}
        B_r = \frac{A_0}{r^5 \sin\theta} \frac{180\degr}{\Delta a} \cos \left[ \frac{180\degr (\lambda+11.5\degr)}{2\Delta a} \right] \sin \left[ \frac{180\degr (\lambda+11.5\degr)}{2\Delta a} \right],\\
        B_\theta = \frac{3A_0}{r^5 \sin\theta} \cos^2 \left[ \frac{180\degr (\lambda+11.5\degr)}{2\Delta a}\right],
        \end{eqnarray}
        
        \noindent where $\Delta a = 37.2 \degr$ represents half the width of the arcade system, $\lambda = 90\degr-\theta$ the solar latitude, $A_0 = 0.73\,\mathrm{G \cdot R_{\sun}^5}$, and the whole arcade system is shifted by $11.5\degr$ to the south. These components were only added to the dipole magnetic field in the region determined by the latitude $ \lambda \in [-48.7\degr, 25.8\degr]$. This configuration provides a polar magnetic field strength of 1.8 G (or ${1.8 \times 10^5}$ nT), and a maximum arcade strength of 1.57 G  (or ${1.57 \times 10^5}$ nT), measured at the first cell of the domain. \\
        Once the solar wind has reached a steady-state solution after $\approx200$ h, CMEs are obtained by applying time-dependent shearing motions at the inner boundary, thus magnetically stressing the corona above the selected region. The shear was introduced through the following additional azimuthal component of the speed:  
        
        \begin{equation}
        v_\phi = v_0(\alpha^2-\Delta \theta^2)^2 \sin\alpha \sin[180\degr(t-t_0)/\Delta t],  
        \end{equation}
        
        \noindent where $\alpha=\lambda - \lambda_0$. This flow is almost symmetric with respect to the latitude of the southernmost polarity inversion line, which is approx. $\lambda_0=-41\degr$, and spans over $2 \Delta \theta=17.2\degr$. It is applied for $\Delta t=16$ h starting from $t_0=0$ h and it has a slow increase and decrease, with a maximum at half the time interval. Throughout the simulations, the scaling factor $v_0$ is given such that $v_\phi$ does not exceed $10\%$ of the local Alfv\'{e}n speed. A representation of the shearing profile can be seen in Figure \ref{fig:shear}.       
        
        \section{Results} \label{sec:results}    
        
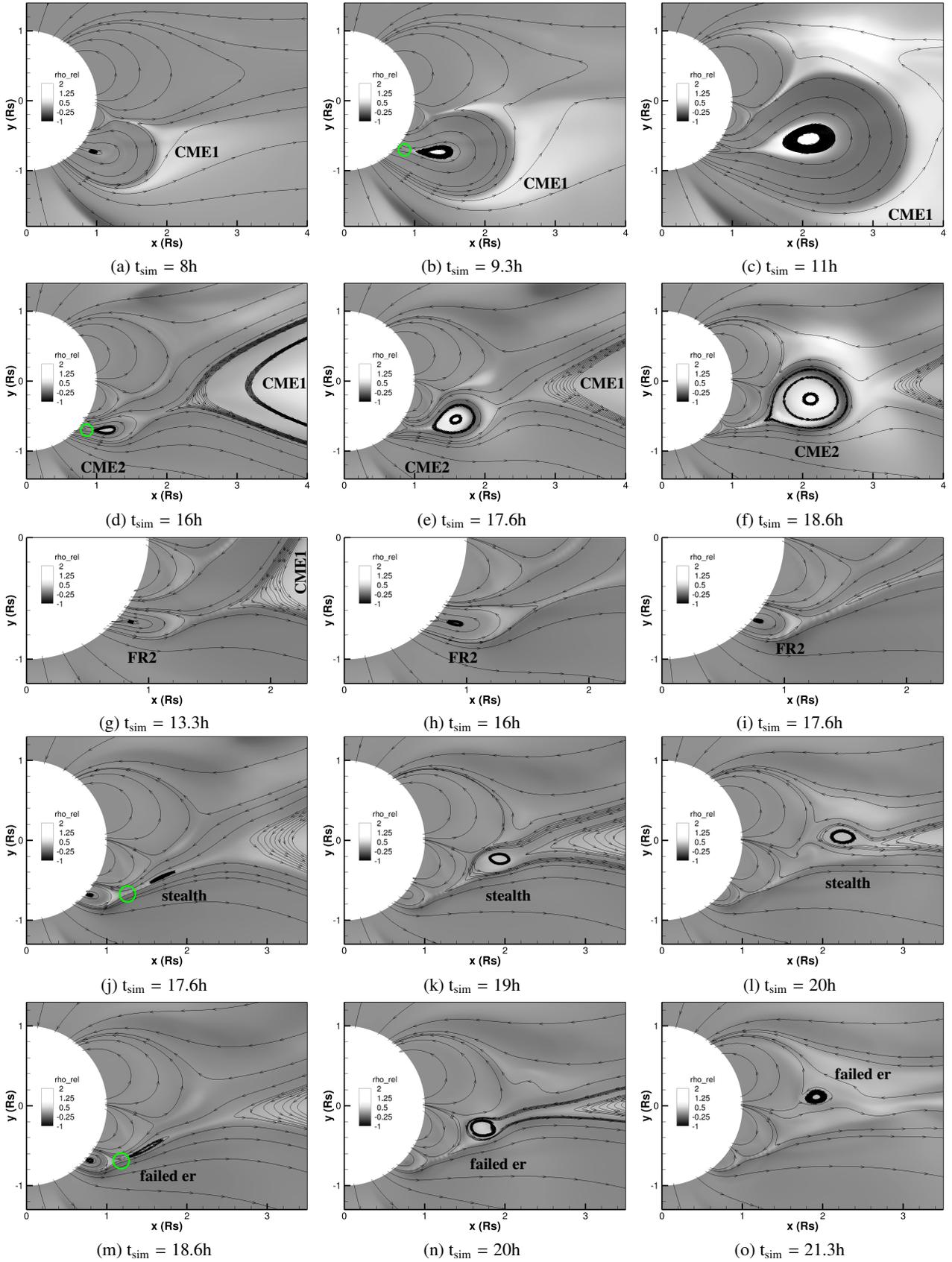
\begin{figure*}    
        \centering
        
        \subfloat[$\mathrm{t_{sim}=8h}$ \label{fig:cme1_1}]{
                        \begin{overpic}[width=0.3\linewidth]{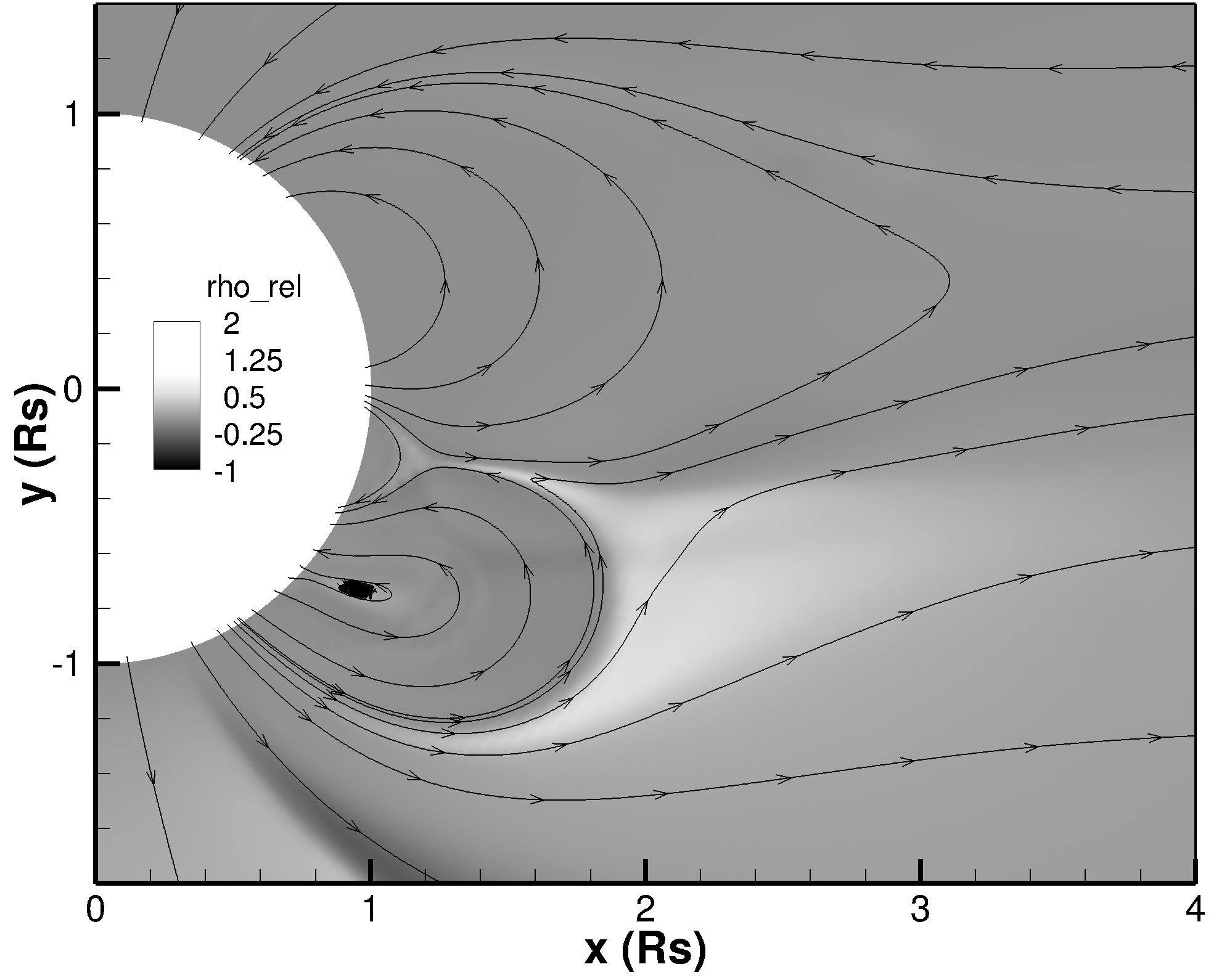}
                                \put(55,30){\tiny \textbf{CME1}}
                        \end{overpic}}
        \subfloat[$\mathrm{t_{sim}=9.3 h}$ \label{fig:cme1_2}]{
                \begin{overpic}[width=0.3\linewidth]{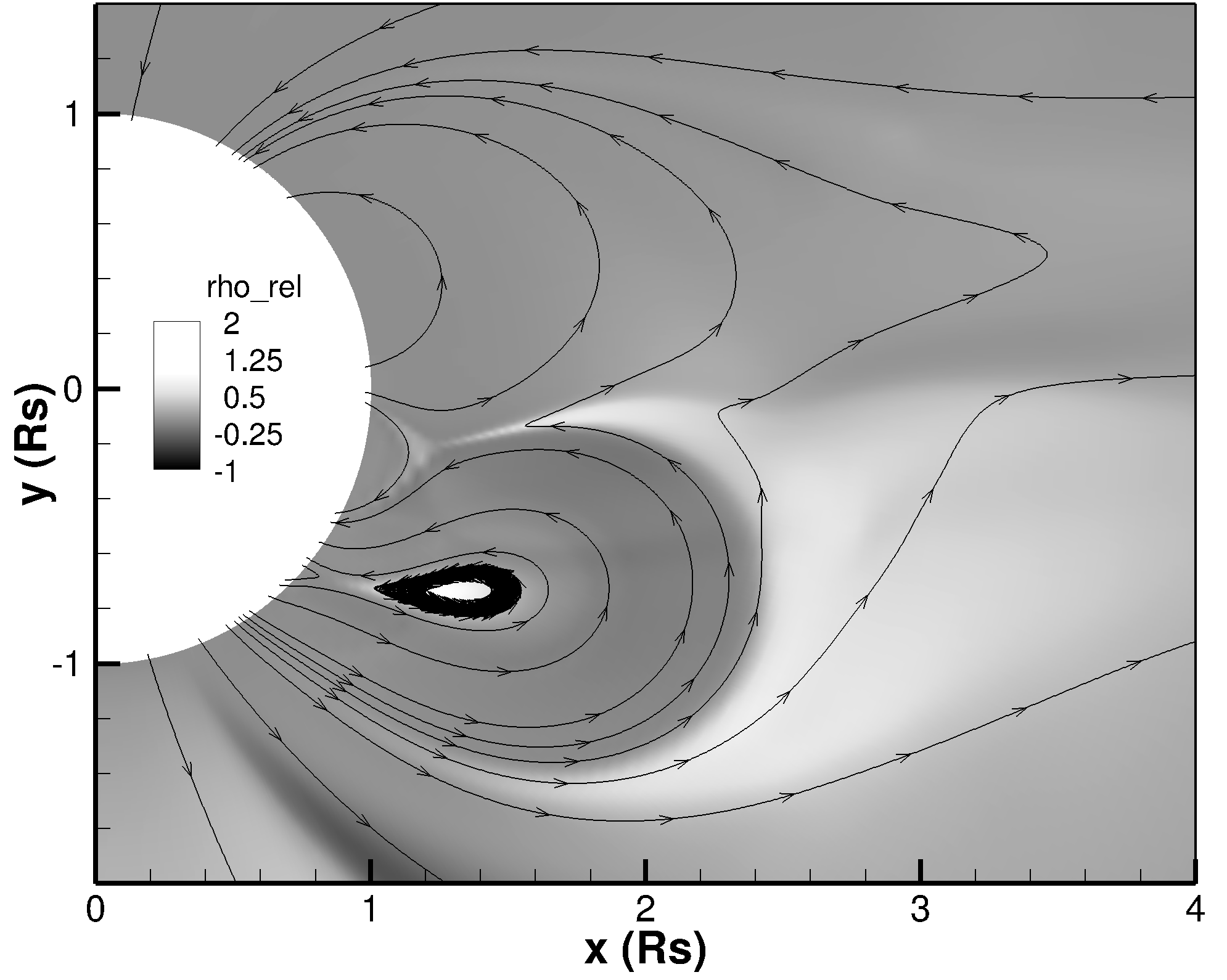}
                        \put(65,20){\tiny \textbf{CME1}}
                        \thicklines
                        \put(27,32.4){\color{green} \circle{4}}
                \end{overpic}}
        \subfloat[$\mathrm{t_{sim}=11 h}$ \label{fig:cme1_3}]{
                \begin{overpic}[width=0.3\linewidth]{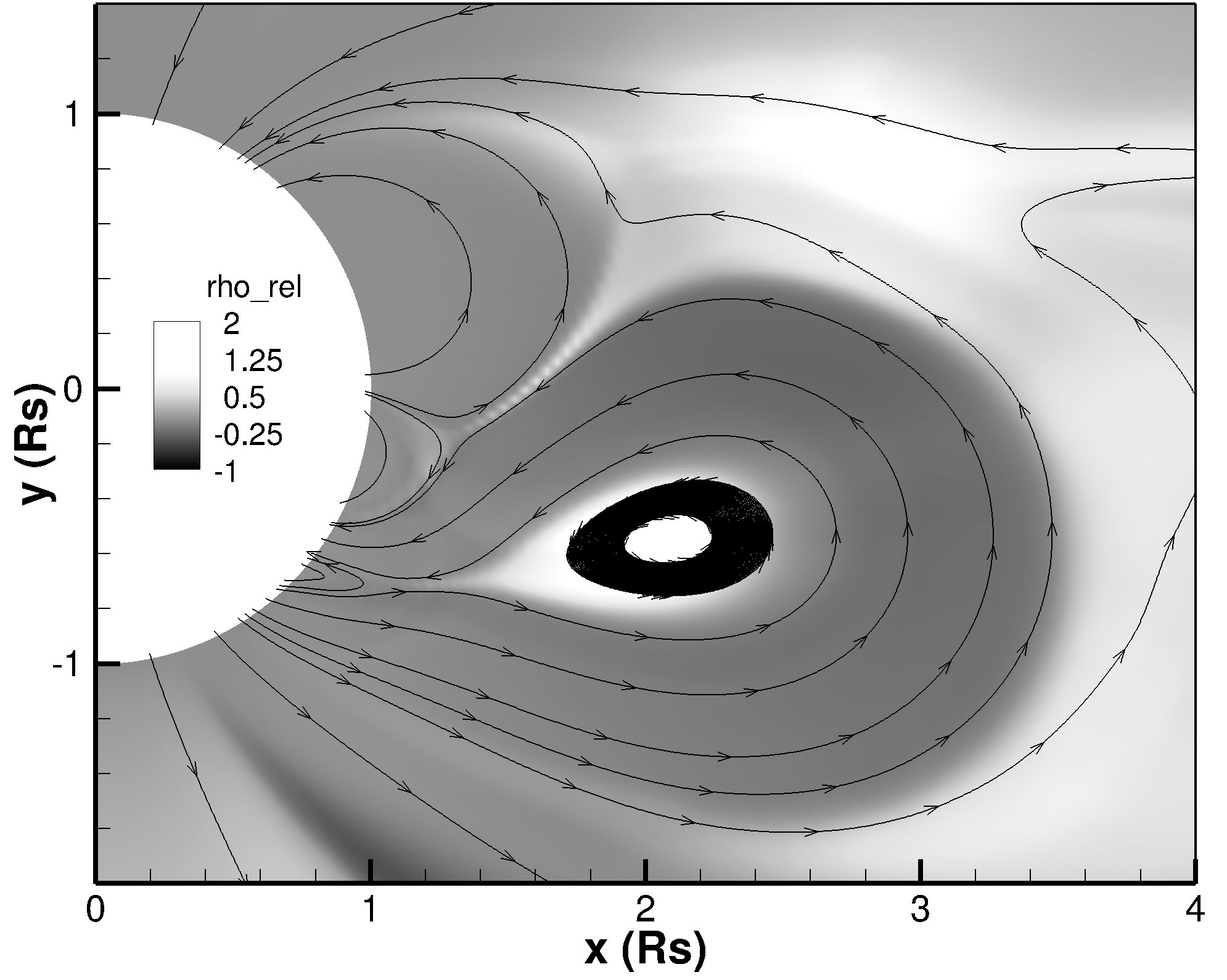}
                        \put(80,10){\tiny \textbf{CME1}}
                \end{overpic}}\\        
        \vspace{-0.3cm}

        \subfloat[$\mathrm{t_{sim}=16 h}$ \label{fig:double_1}]{
                \begin{overpic}[width=0.3\linewidth]{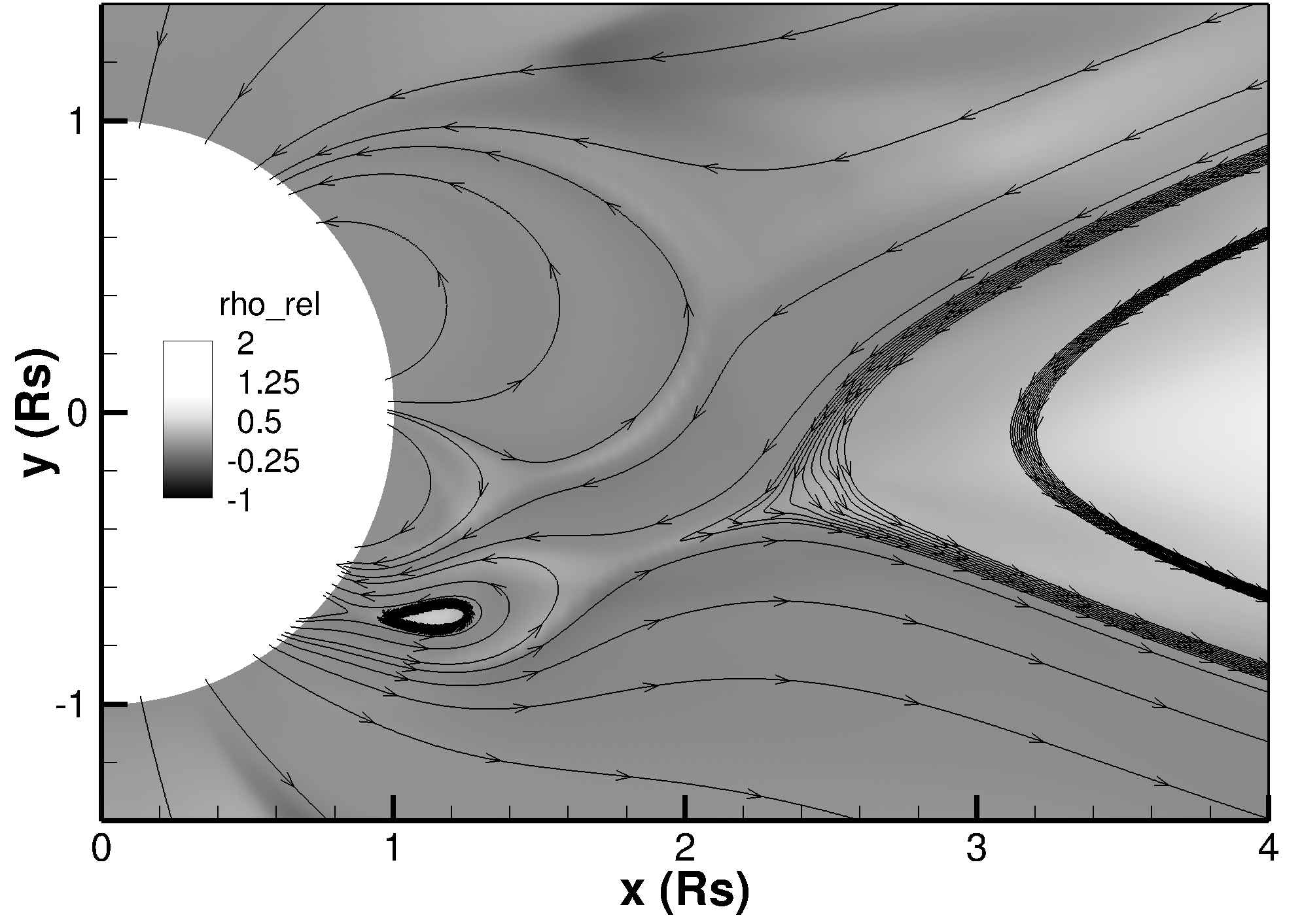}
                        \put(83,37){\tiny \textbf{CME1}}                        
                        \put(25,10){\tiny \textbf{CME2}}
                        \thicklines
                        \put(27,23.5){\color{green} \circle{4}}                         
                \end{overpic}}
        \subfloat[$\mathrm{t_{sim}= 17.6 h}$ \label{fig:double_2}]{
                \begin{overpic}[width=0.3\linewidth]{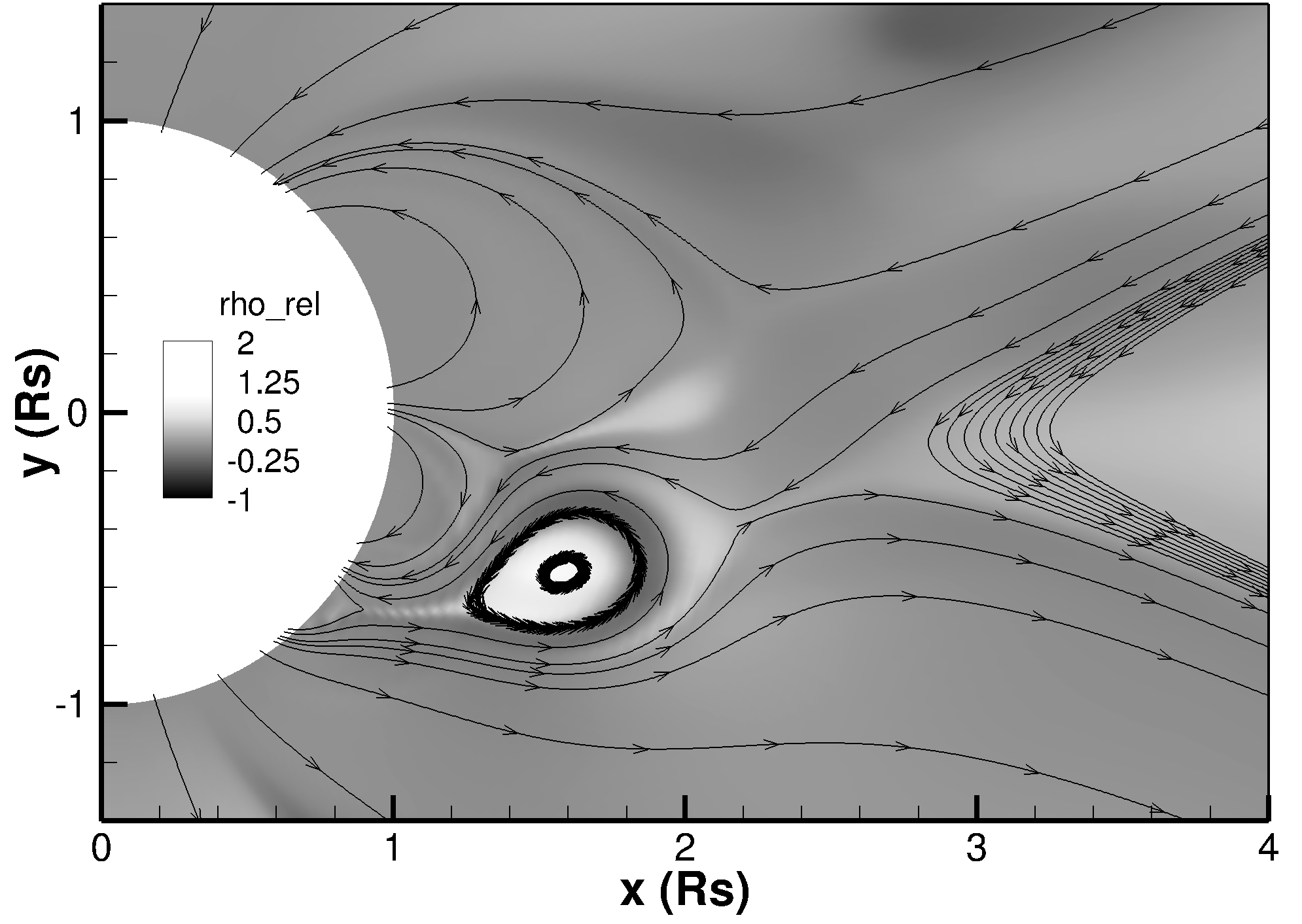}
                        \put(83,37){\tiny \textbf{CME1}}                        
                        \put(27,10){\tiny \textbf{CME2}}                                                 
                \end{overpic}}
        \subfloat[$\mathrm{t_{sim}=18.6 h}$ \label{fig:double_3}]{
                \begin{overpic}[width=0.3\linewidth]{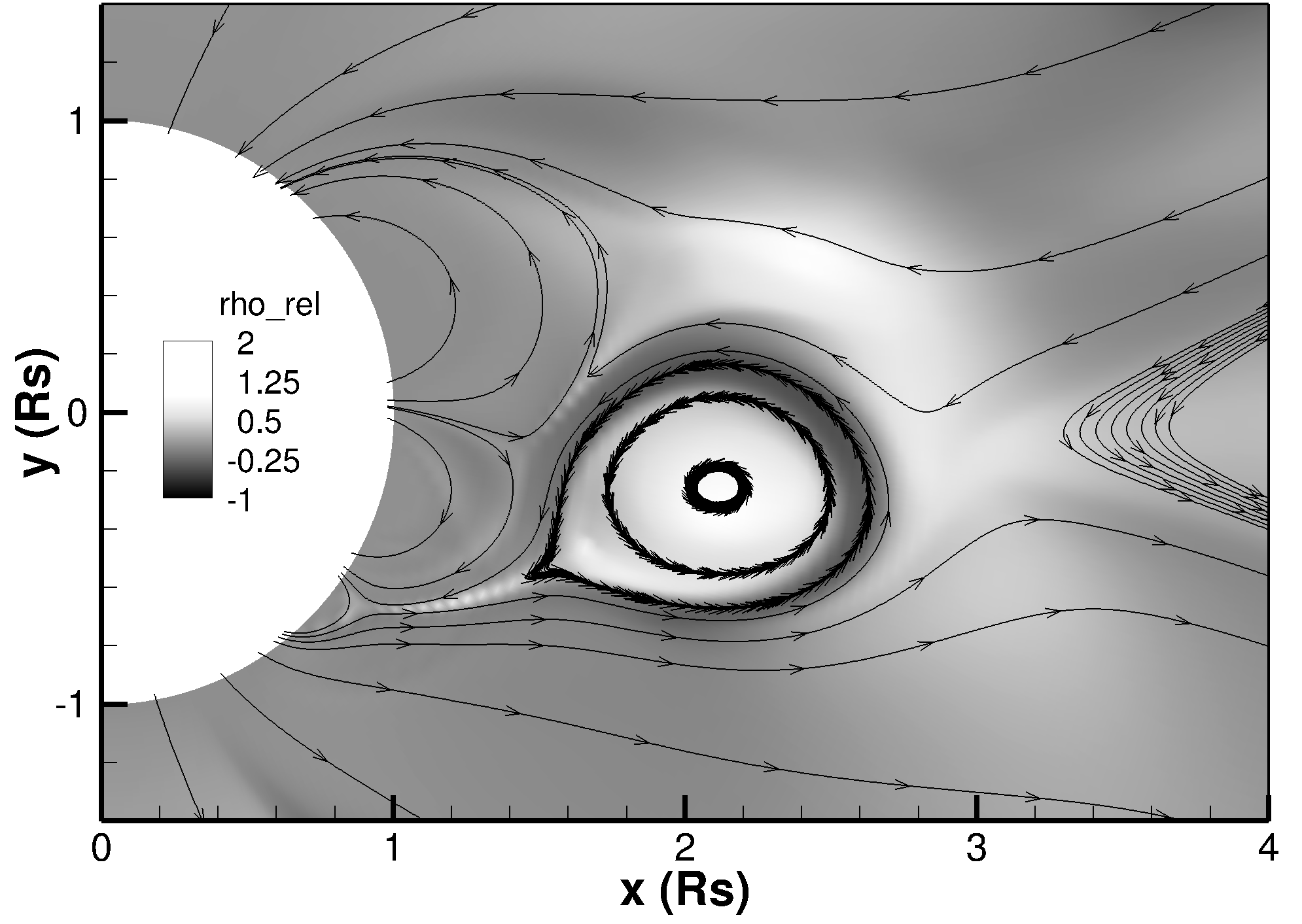}
                        \put(50,15){\tiny \textbf{CME2}}
                \end{overpic}}\\
        \vspace{-0.3cm}         

        \subfloat[$\mathrm{t_{sim}= 13.3 h}$ \label{fig:fr2_1}]{
                \begin{overpic}[width=0.3\linewidth]{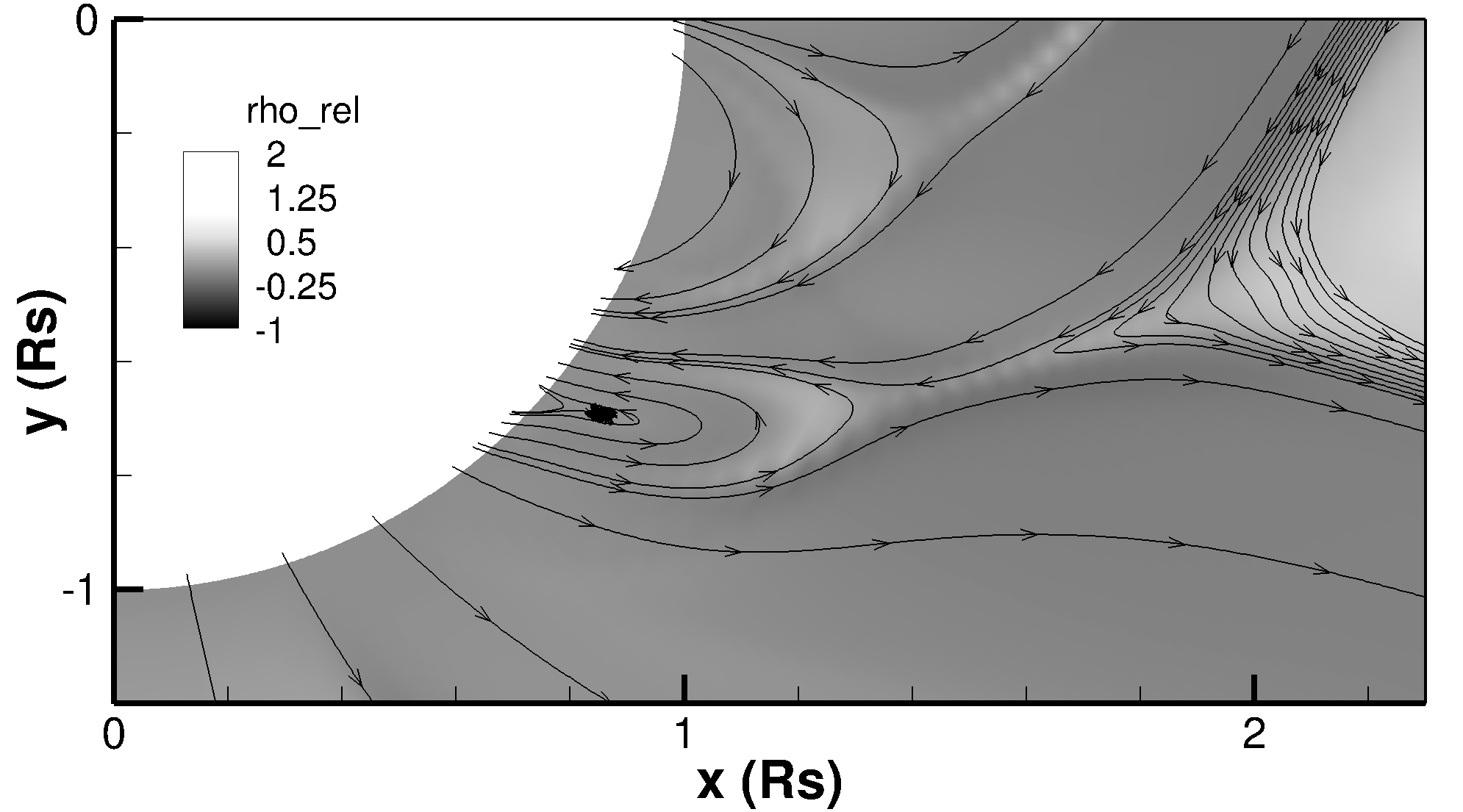}
                        \put(93.5,36){\rotatebox{90}{\tiny \textbf{CME1}}}                         
                        \put(40,14){\tiny \textbf{FR2}}
                \end{overpic}}
        \subfloat[$\mathrm{t_{sim}= 16 h}$ \label{fig:fr2_2}]{
                \begin{overpic}[width=0.3\linewidth]{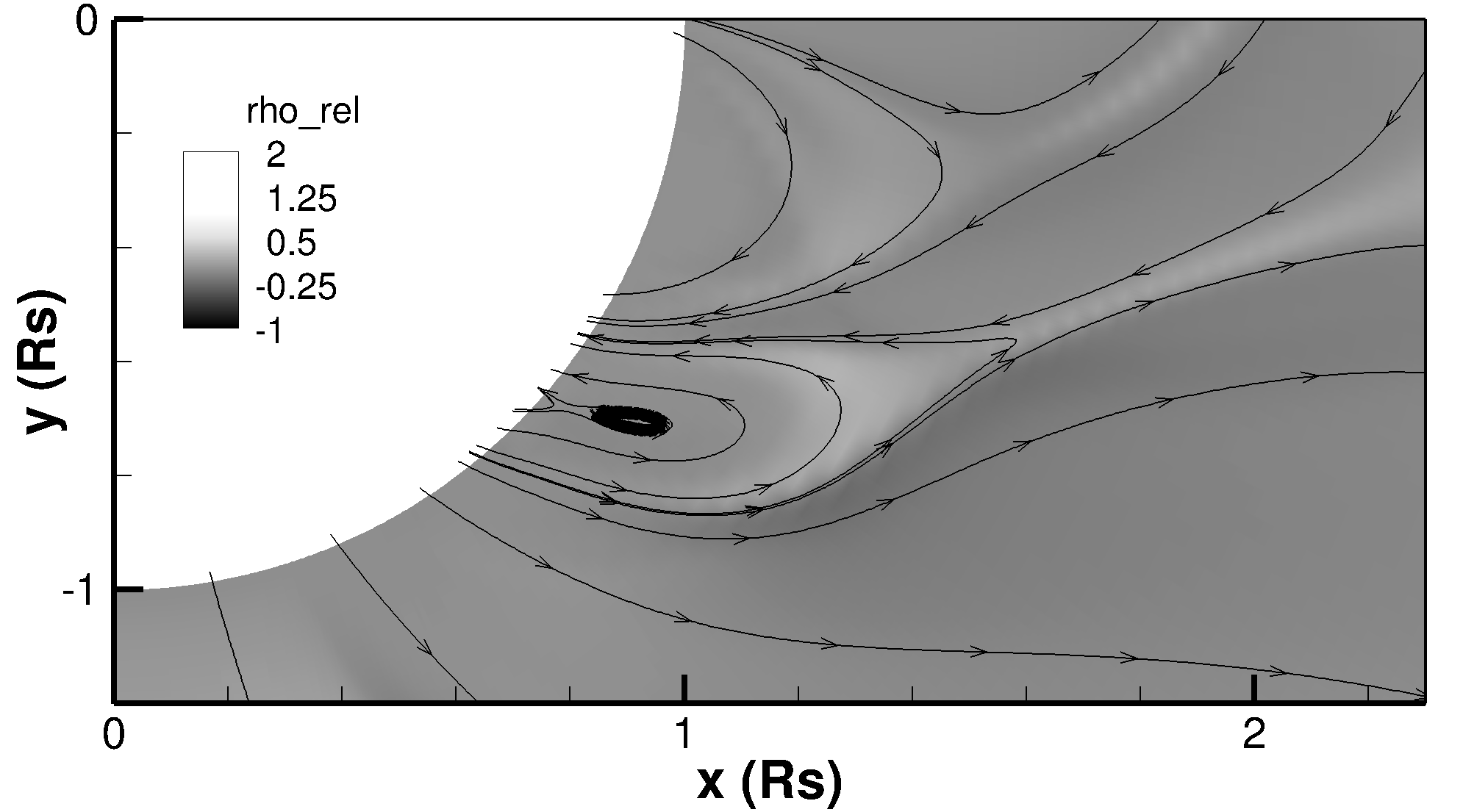}
                        \put(41,14){\tiny \textbf{FR2}}
                \end{overpic}}                  
        \subfloat[$\mathrm{t_{sim}= 17.6 h}$ \label{fig:fr2_3}]{
                \begin{overpic}[width=0.3\linewidth]{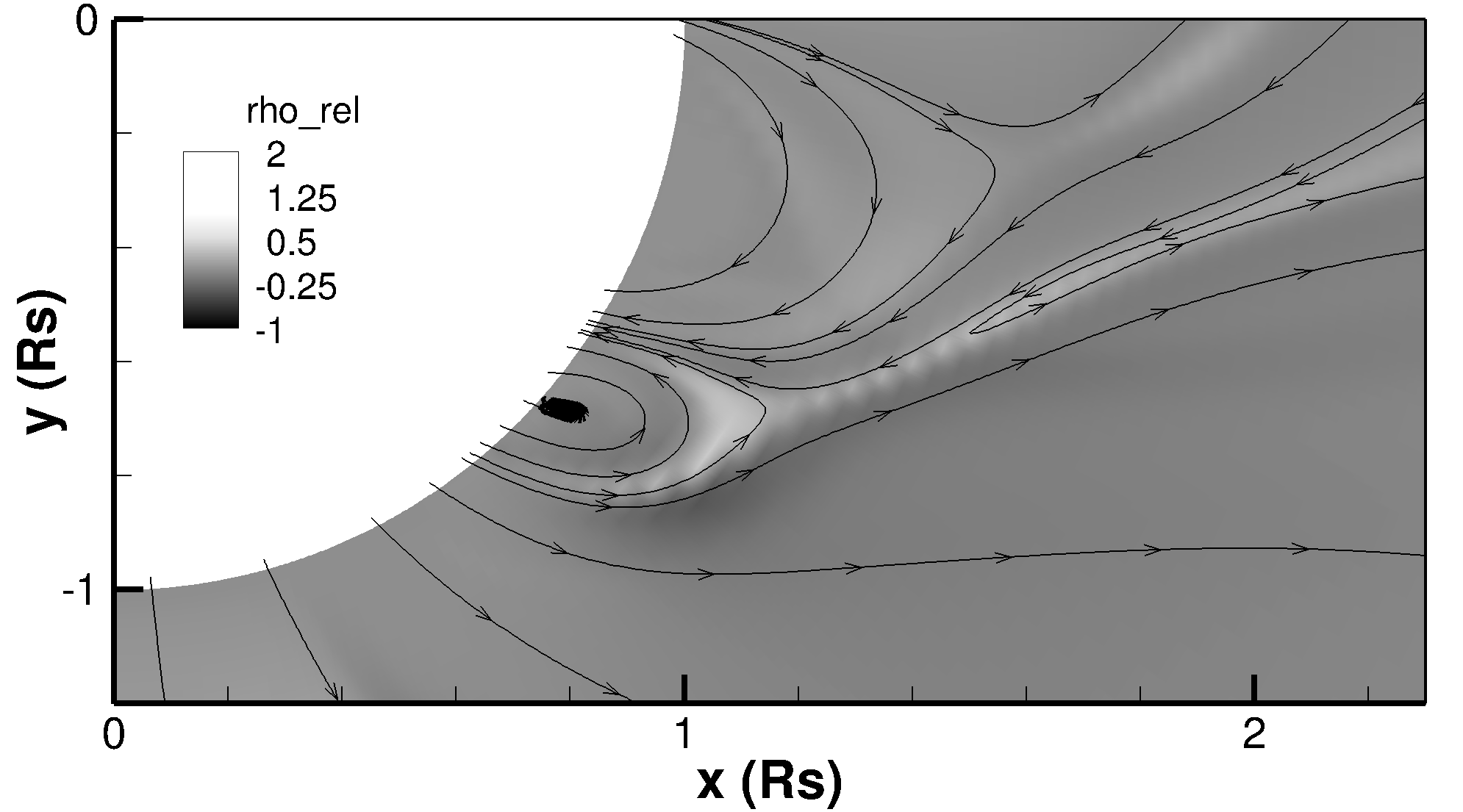}
                        \put(44,17){\tiny \textbf{FR2}}
                \end{overpic}}\\
        \vspace{-0.3cm}

        \subfloat[$\mathrm{t_{sim}=17.6 h}$ \label{fig:stealth_1}]{
                \begin{overpic}[width=0.3\linewidth]{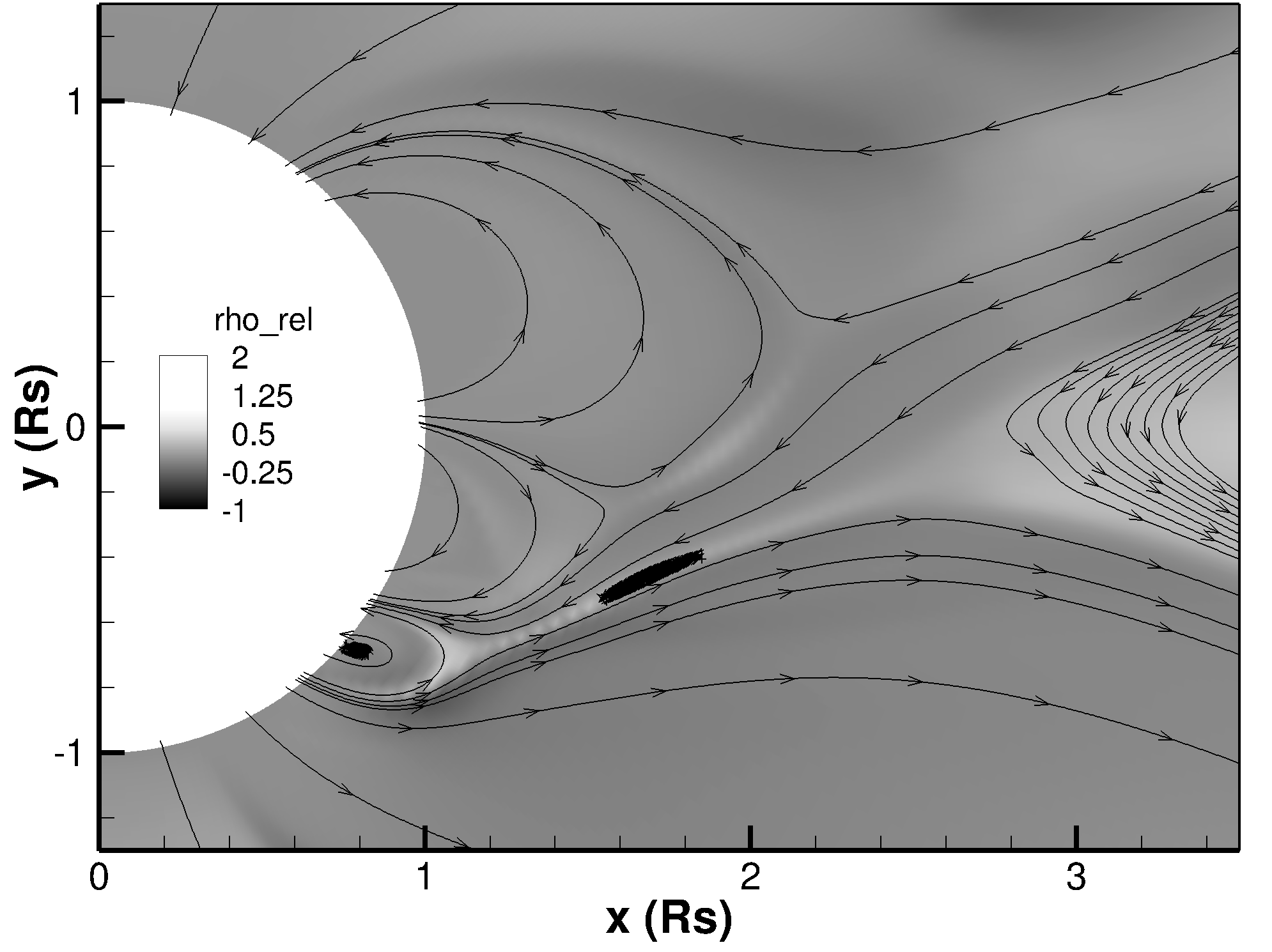}
                        \put(51,22){\tiny \textbf{stealth}}
                        \thicklines
                        \put(40,24){\color{green} \circle{5}}                           
                \end{overpic}}
        \subfloat[$\mathrm{t_{sim}=19 h}$ \label{fig:stealth_2}]{
                \begin{overpic}[width=0.3\linewidth]{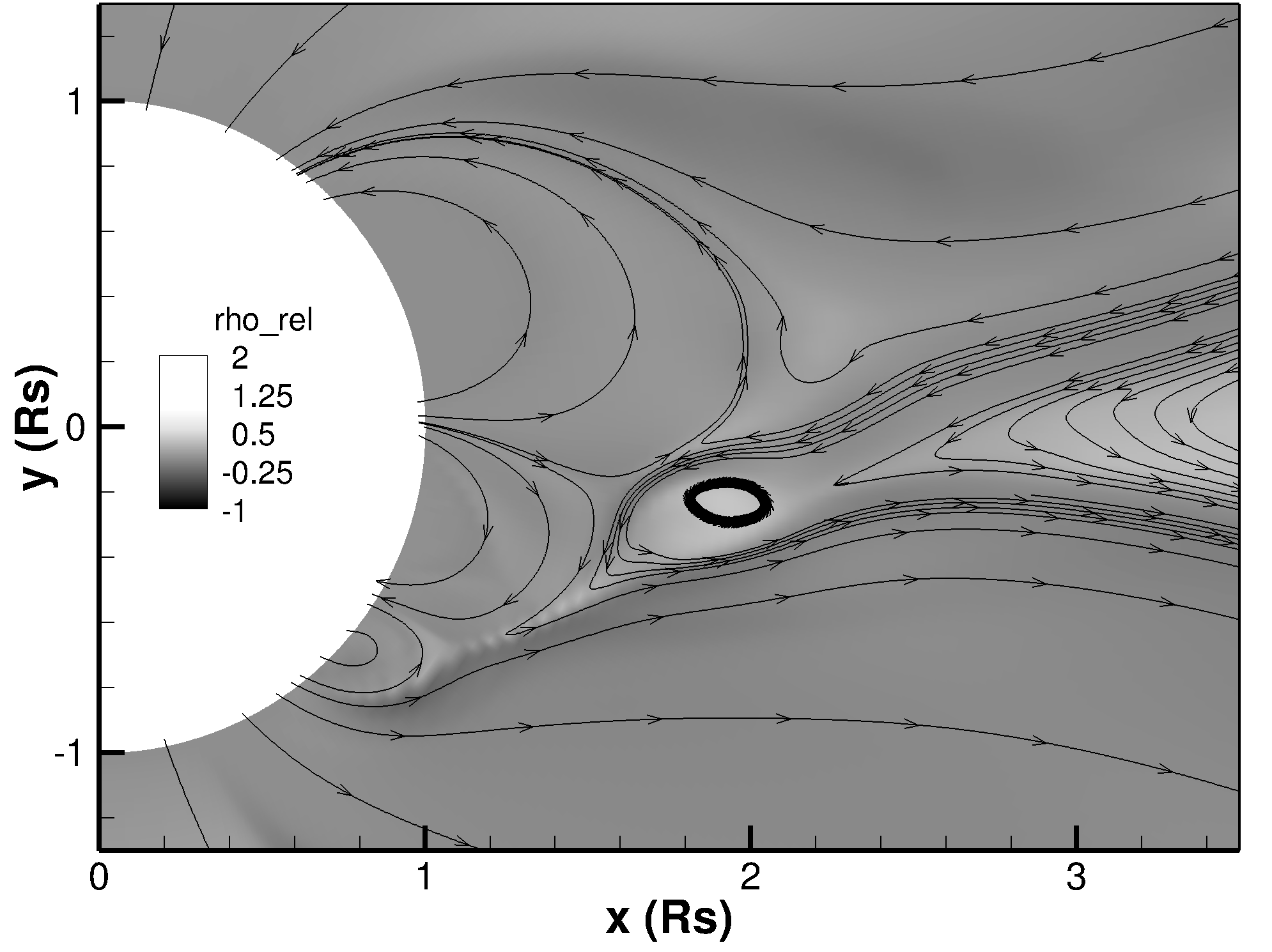}
                        \put(53,22){\tiny \textbf{stealth}}
                \end{overpic}}
        \subfloat[$\mathrm{t_{sim}=20 h}$ \label{fig:stealth_3}]{
                \begin{overpic}[width=0.3\linewidth]{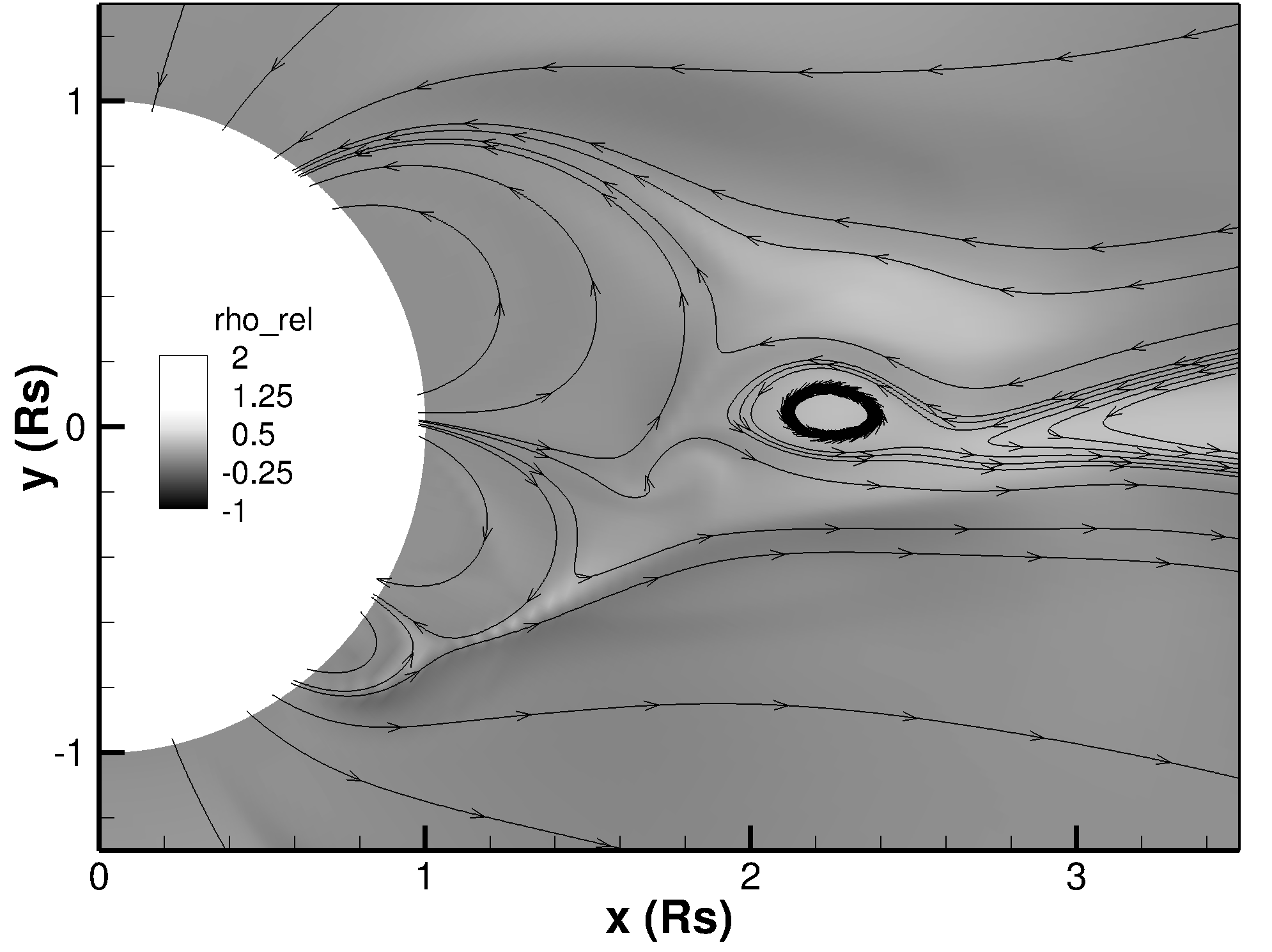}
                        \put(60,26){\tiny \textbf{stealth}}
                \end{overpic}}\\
        \vspace{-0.3cm}

        \subfloat[$\mathrm{t_{sim}=18.6 h}$ \label{fig:failed_1}]{
                \begin{overpic}[width=0.3\linewidth]{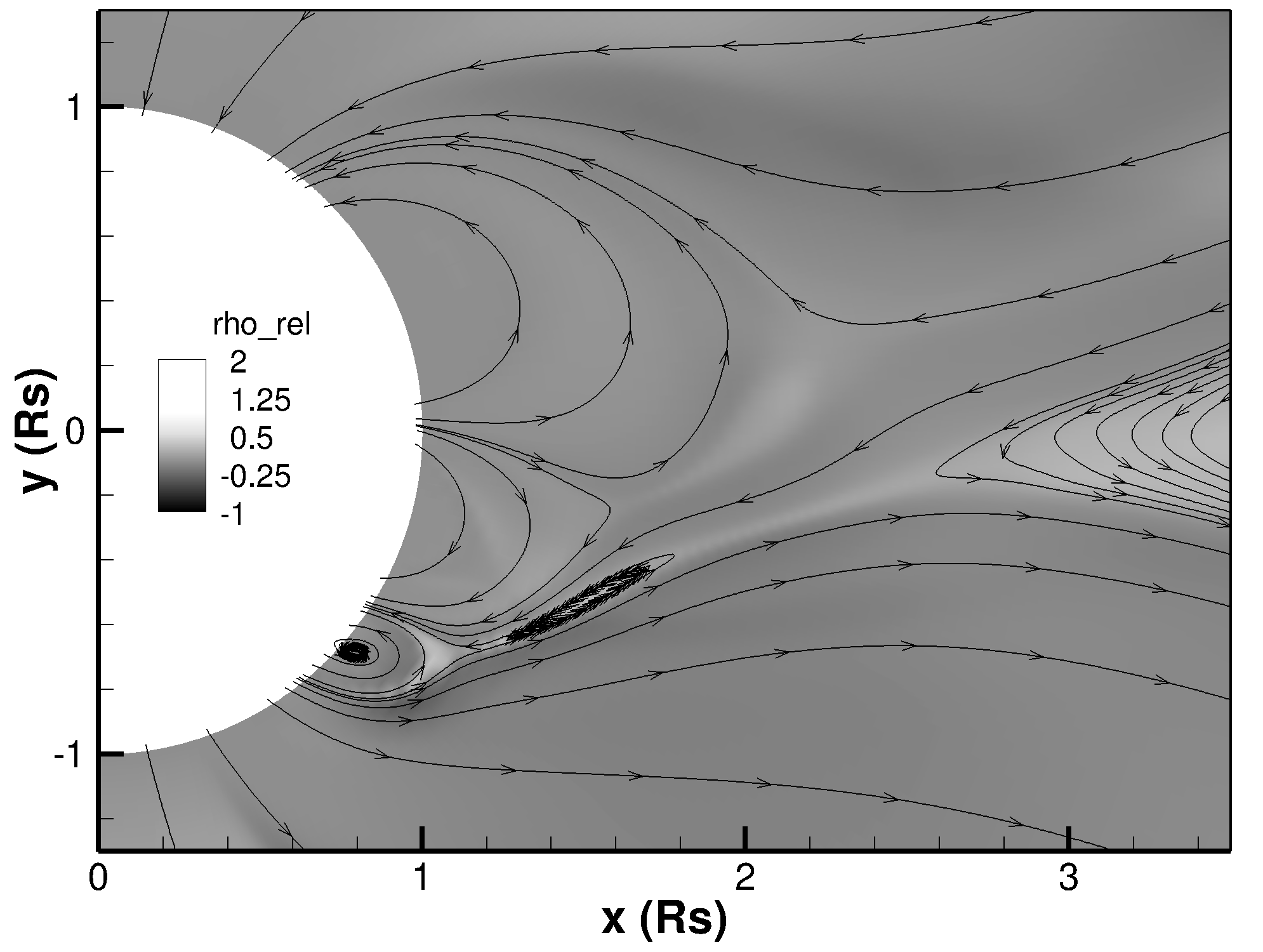}
                        \put(44,17){\tiny \textbf{failed er}}
                        \thicklines
                        \put(38,23.5){\color{green} \circle{5}}                         
                \end{overpic}}
        \subfloat[$\mathrm{t_{sim}=20 h}$ \label{fig:failed_2}]{
                \begin{overpic}[width=0.3\linewidth]{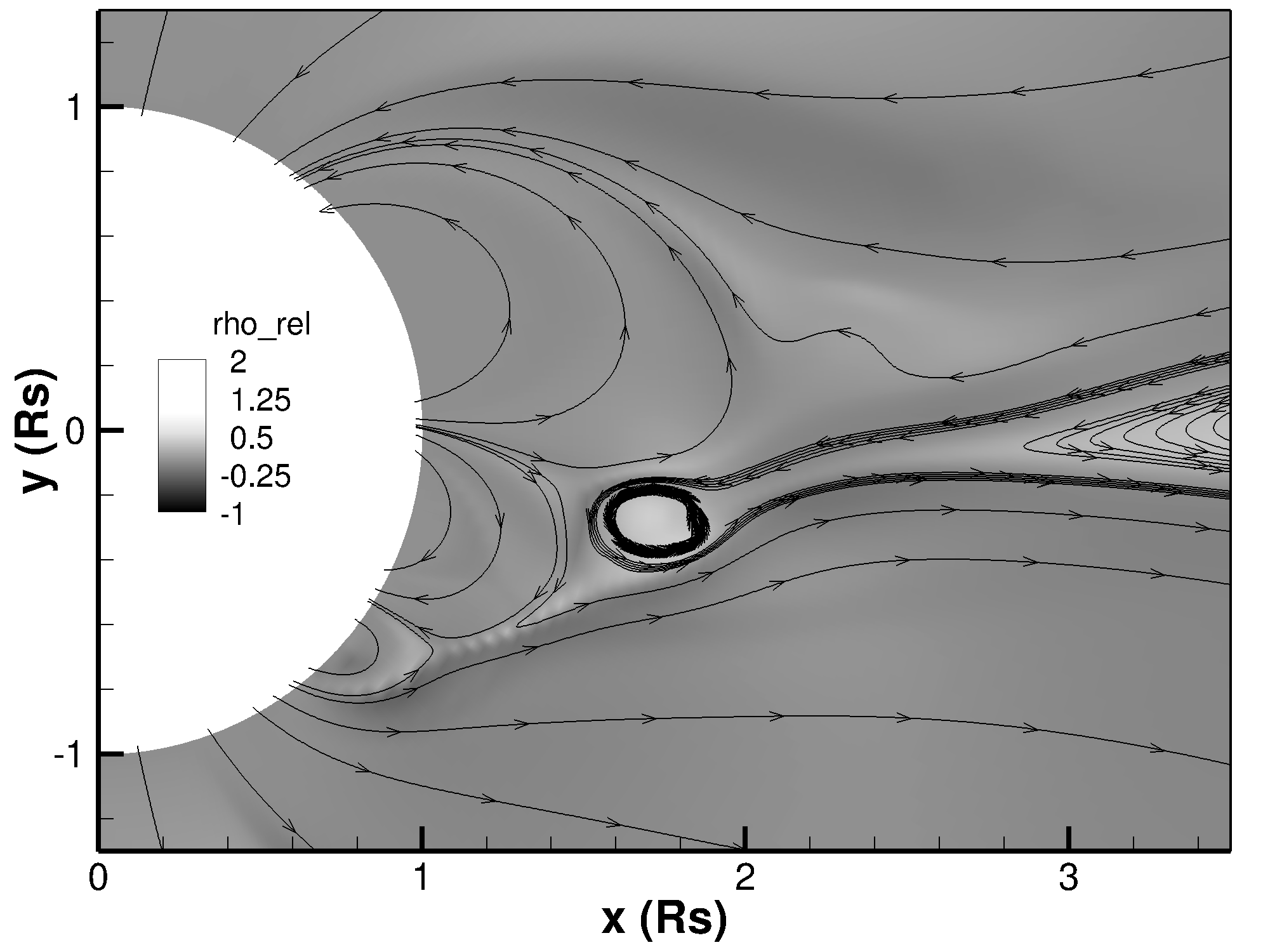}
                        \put(48,21){\tiny \textbf{failed er}}
                \end{overpic}}
        \subfloat[$\mathrm{t_{sim}=21.3 h}$ \label{fig:failed_3}]{
                \begin{overpic}[width=0.3\linewidth]{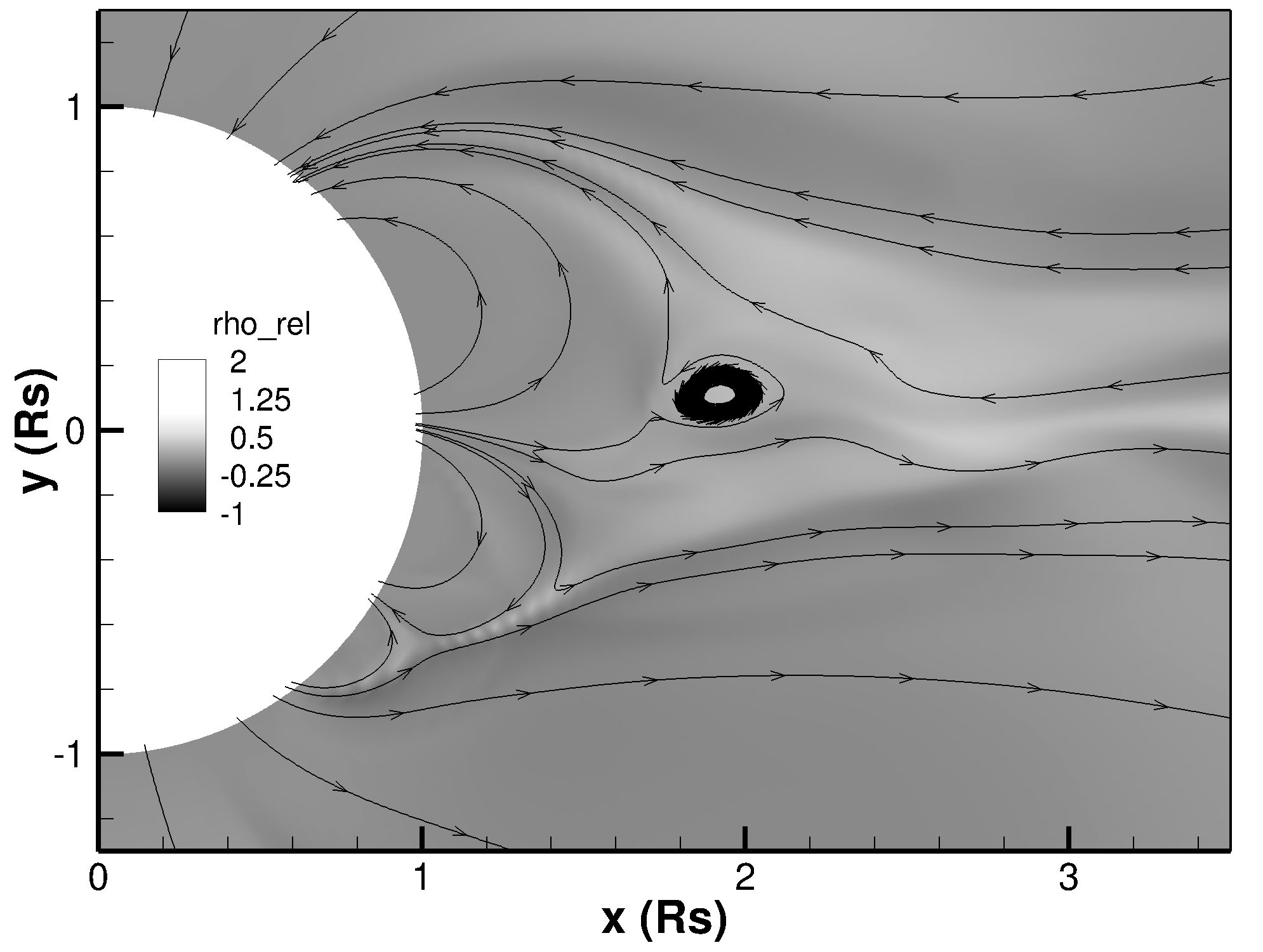}
                        \put(63,50){\tiny \textbf{failed er}}
                \end{overpic}}\\
        
        \caption{Simulated relative density (grey scale) and selected magnetic field lines during the formation phase of flux ropes (FRs), in the case of: first row - CME1; second row - CME2 (double eruption case); third row - second FR (FR2) which falls back to the Sun; fourth row - stealth eruption; fifth row - failed eruption. The green circles indicate reconnection sites. The relative density is $\rho_{rel}=\frac{\rho(t)-\rho_0}{\rho_0}$, where ${\rho_0}$ is the density of the initial relaxed state before the shear.}
        \label{fig:simulations_snapshots}
       \end{figure*}

\begin{figure*}    
        \centering
        \hspace*{0.13in}        
        \subfloat[time: 22 September 2009, 05:39 UT; $\Delta$t = 8h 39min \label{fig:cme1_observations}]{
                \begin{overpic}[width=0.447\linewidth]{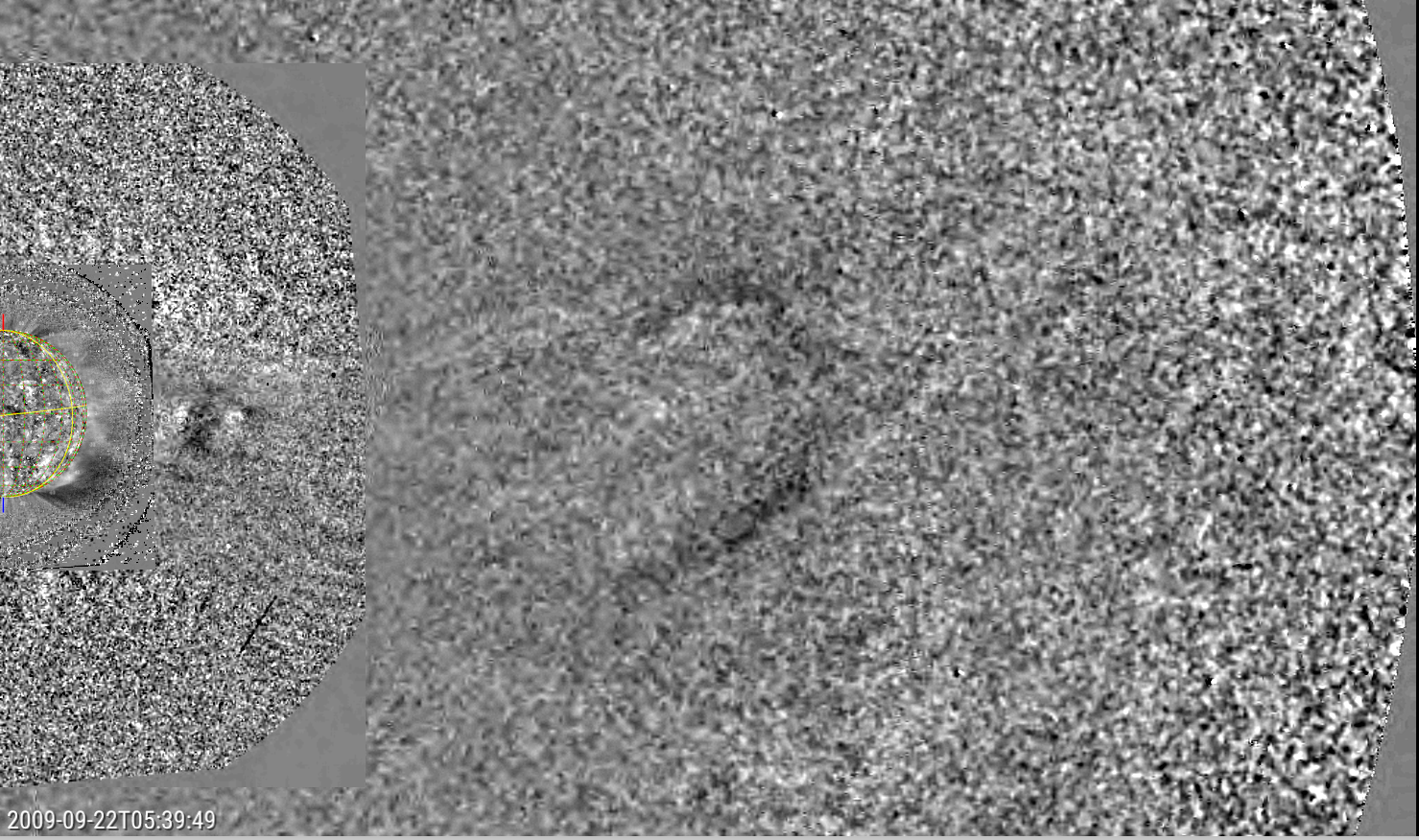}
                        \put(10,37){\colorbox{white}{\parbox{0.043\linewidth}{\tiny \textbf{CME2}}}}
                        \put(53,45){\colorbox{white}{\parbox{0.043\linewidth}{\tiny \textbf{CME1}}}}                        
                \end{overpic}}          
        \hspace*{0.23in}        
        \subfloat[time: 22 September 2009, 08:24 UT; $\Delta$t = 11h 24min \label{fig:cme2_observations}]{
                \begin{overpic}[width=0.447\linewidth]{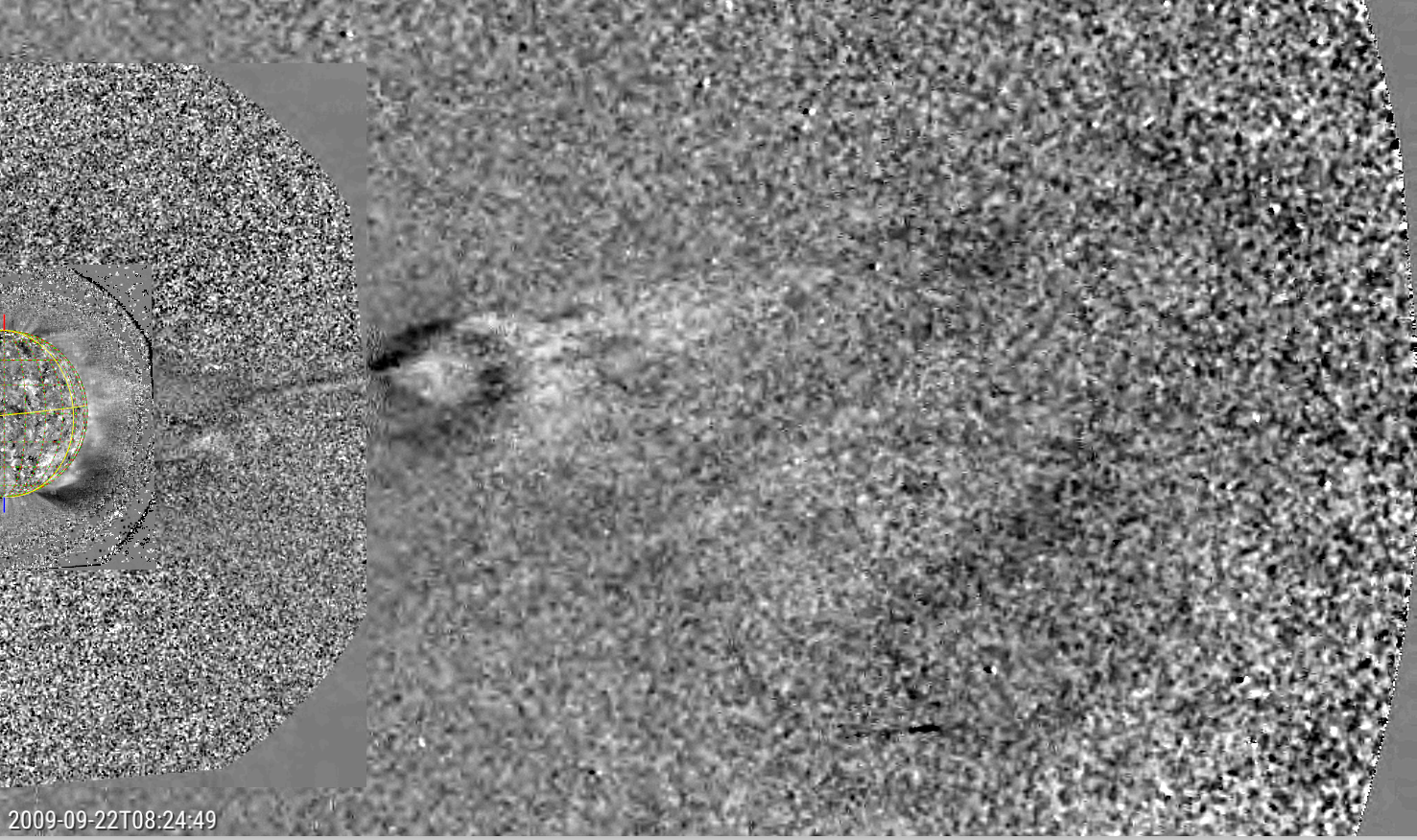}
                        \put(25,43){\colorbox{white}{\parbox{0.043\linewidth}{\tiny \textbf{CME2}}}}
                        \put(70,50){\colorbox{white}{\parbox{0.043\linewidth}{\tiny \textbf{CME1}}}}                        
                \end{overpic}}\\ 
                
        \subfloat[$\mathrm{t_{sim}=17.6 h}$; $\Delta$t = 8 h \label{fig:cme1_sim_de}]{
                \begin{overpic}[width=0.48\linewidth]{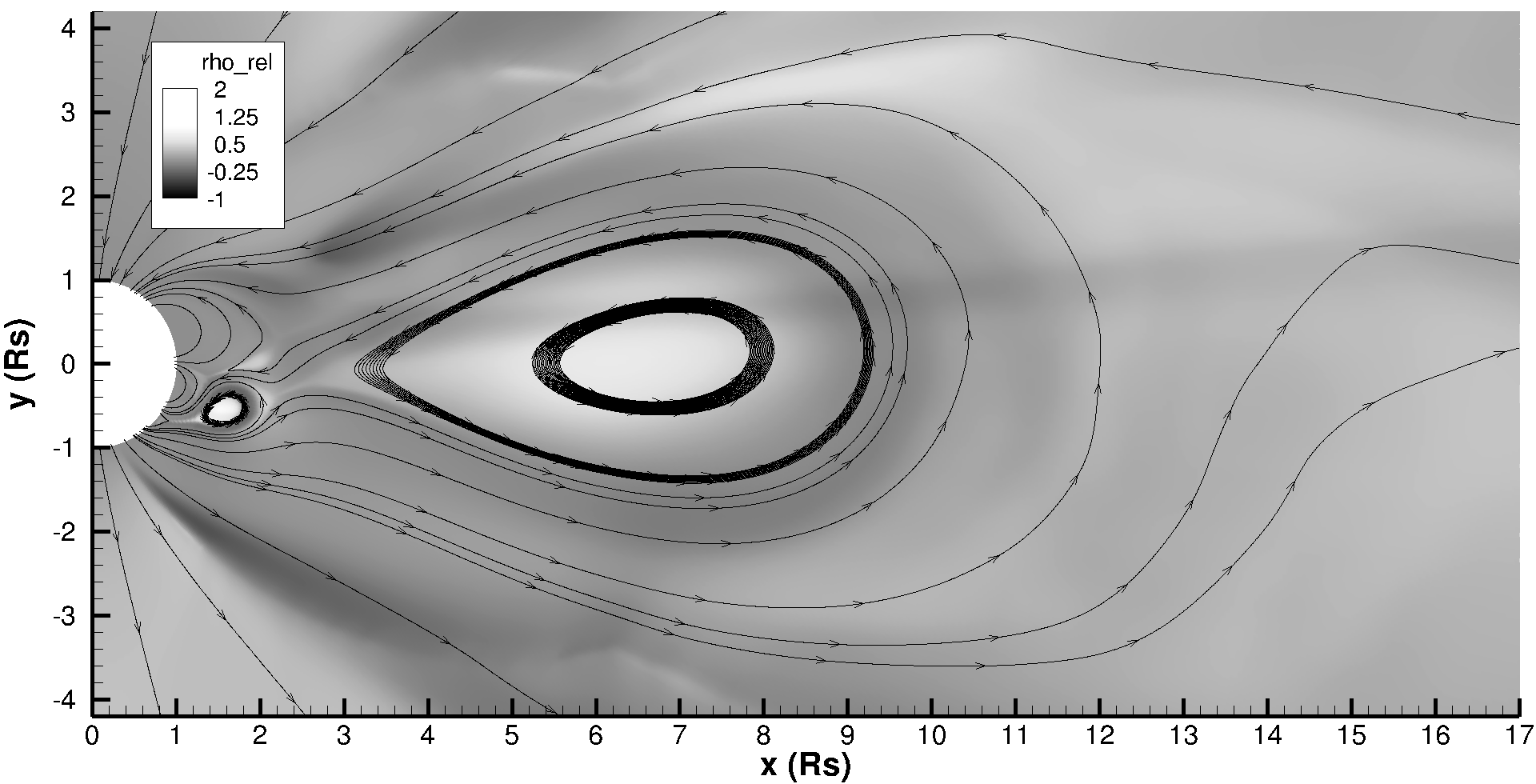}
                        \put(12,18){\tiny \textbf{CME2}}
                        \put(53,40){\tiny \textbf{CME1}}                        
                \end{overpic}}
        \subfloat[$\mathrm{t_{sim}=22 h}$; $\Delta$t = 12h 20min \label{fig:cme2_sim_de}]{
                \begin{overpic}[width=0.48\linewidth]{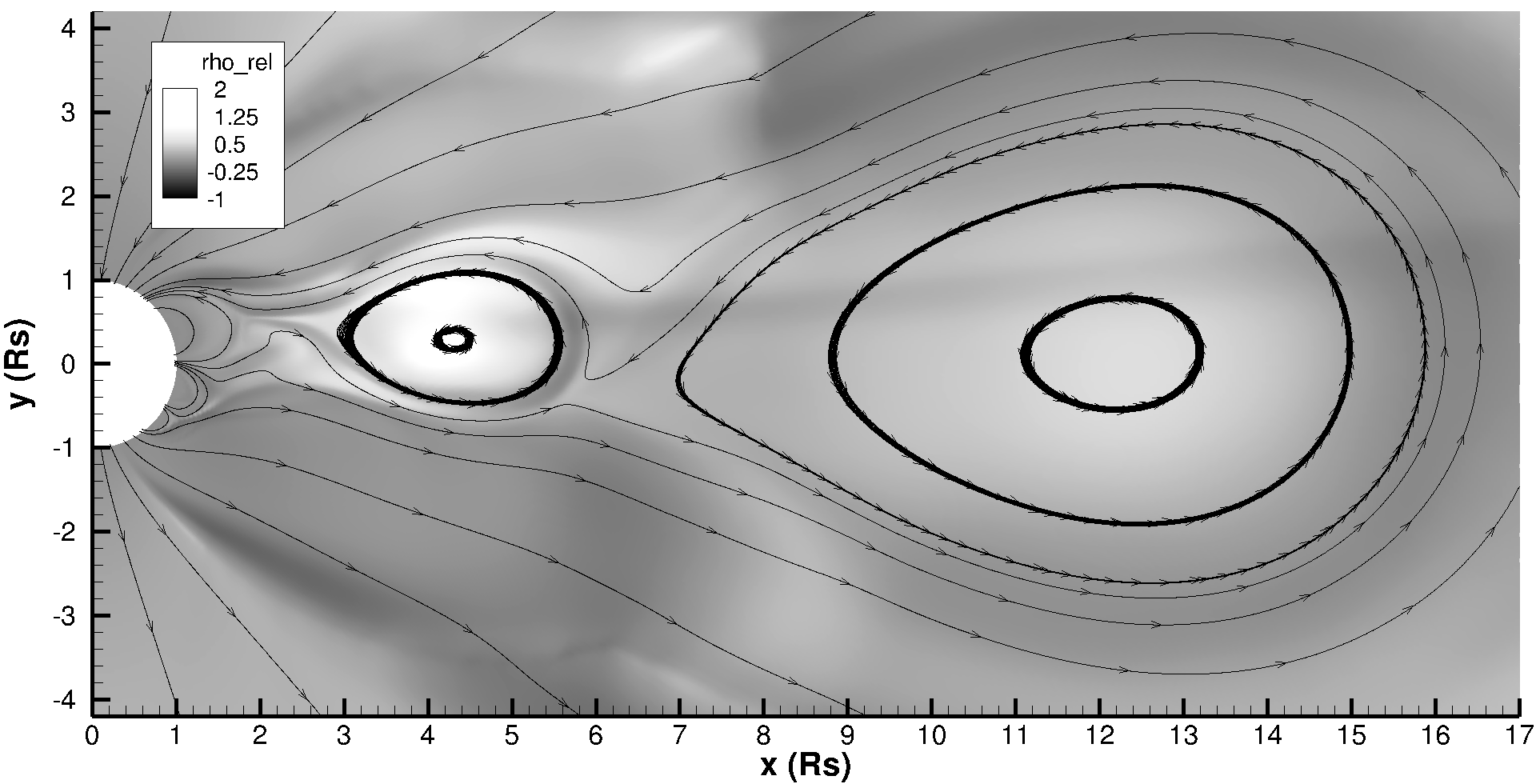}
                        \put(25,18){\tiny \textbf{CME2}}
                        \put(70,33){\tiny \textbf{CME1}}                        
                \end{overpic}}\\

        \subfloat[$\mathrm{t_{sim}=17.6 h}$; $\Delta$t = 8 h \label{fig:cme1_sim_se}]{
                \begin{overpic}[width=0.48\linewidth]{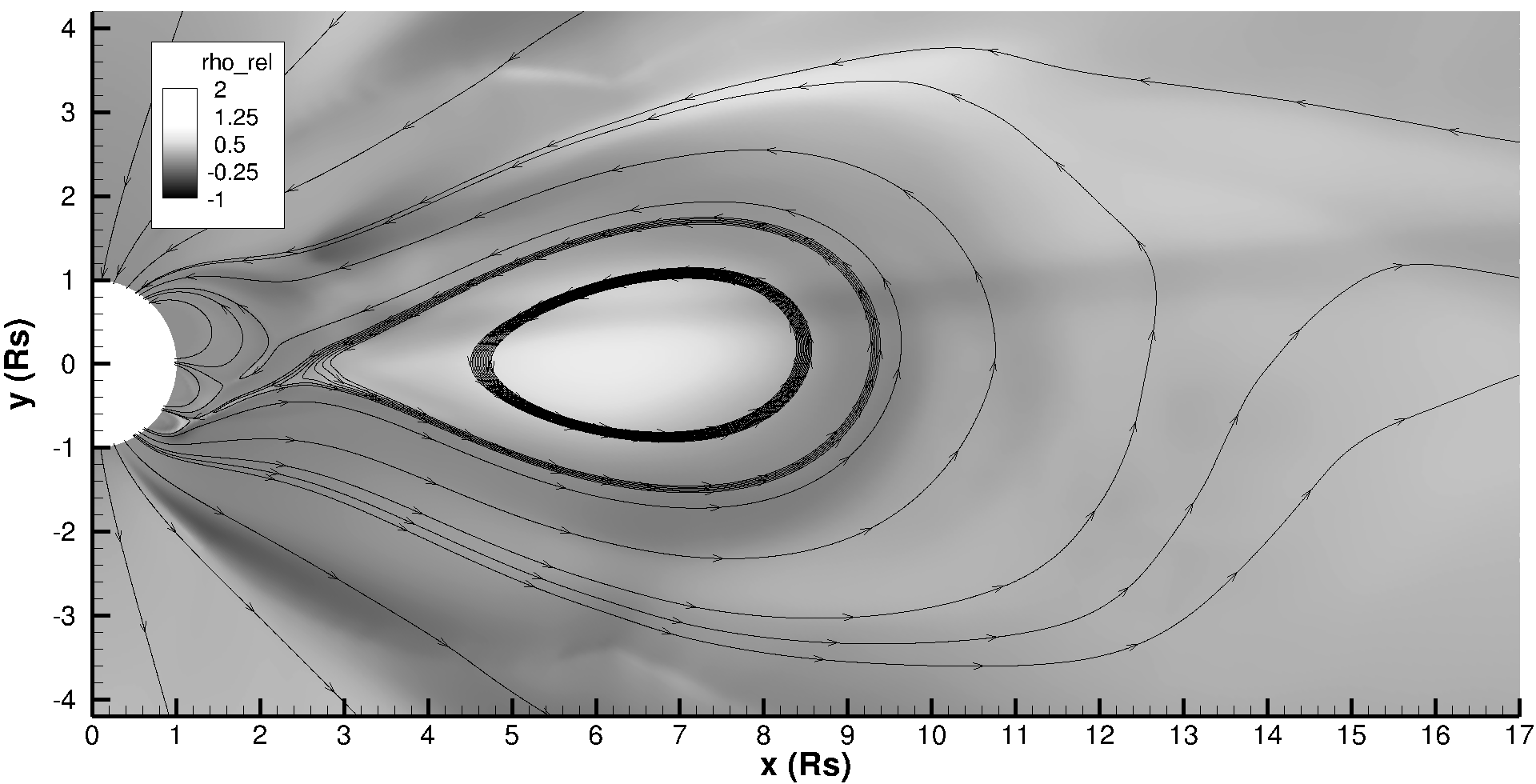}
                        \put(37,27){\tiny \textbf{CME1}}                        
                \end{overpic}}
        \subfloat[$\mathrm{t_{sim}=25 h}$; $\Delta$t = 15h 20min \label{fig:cme2_sim_se}]{
                \begin{overpic}[width=0.48\linewidth]{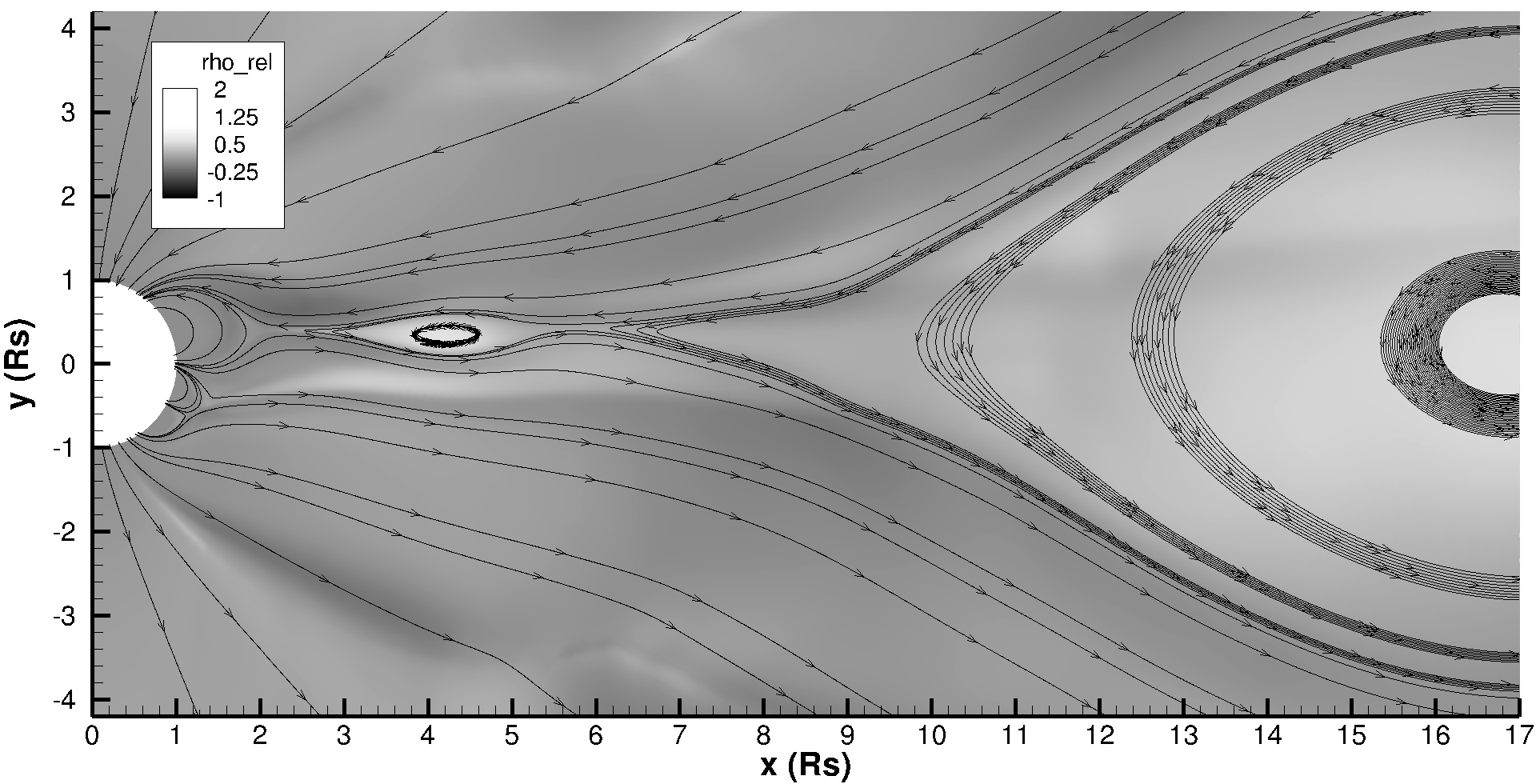}
                        \put(25,25){\tiny \textbf{stealth}}
                        \put(80,28){\tiny \textbf{CME1}}                        
                \end{overpic}}\\
                
        \subfloat[$\mathrm{t_{sim}=17.6 h}$; $\Delta$t = 8 h \label{fig:cme1_sim_fe}]{
                \begin{overpic}[width=0.48\linewidth]{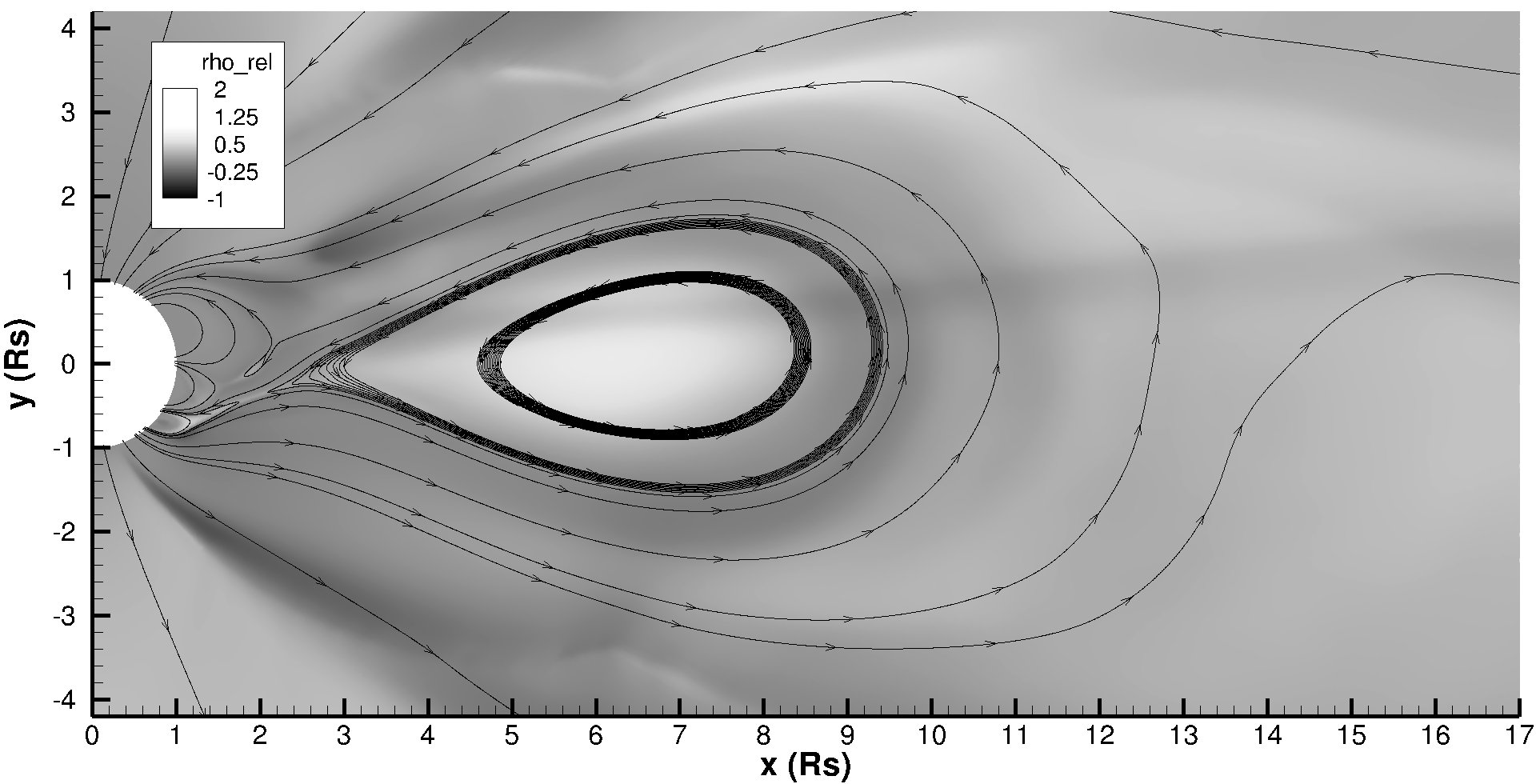}
                        \put(37,27){\tiny \textbf{CME1}}                        
                \end{overpic}}
        \subfloat[$\mathrm{t_{sim}=20.3 h}$; $\Delta$t = 10h 40min \label{fig:cme2_sim_fe}]{
                \begin{overpic}[width=0.48\linewidth]{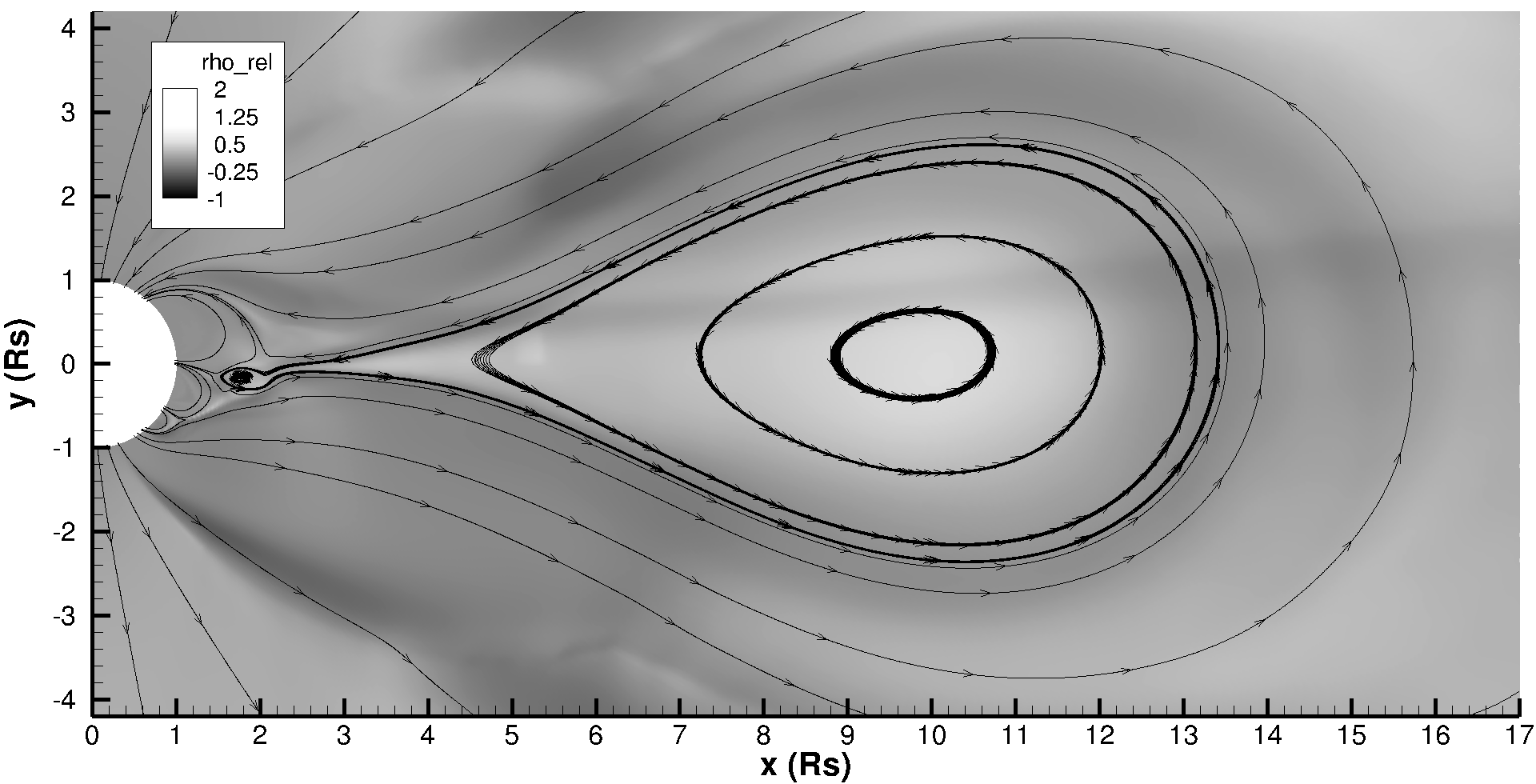}
                        \put(12,20){\tiny \textbf{failed er}}                           
                        \put(56,32){\tiny \textbf{CME1}}                        
                \end{overpic}}\\ 
                
        \caption{First column: Snapshots of observations and MPI-AMRVAC simulations taken when CME1 front is located at 10 R$_{\sun}$. Second column: Snapshots of observations and simulations taken when CME2 front is located at 6 R$_{\sun}$ (except (h)). Panels (a) and (b): Running difference images taken from EUVI, COR1, and COR2 instruments on board STEREO-B and visualised in JHelioviewer software. Snapshots of the simulated relative density and selected magnetic field lines: panels (c) and (d) indicate the double eruption scenario, (e) and (f) show the stealth eruption scenario, and (g) and (h) indicate the failed eruption scenario. The parameter $\Delta$t represents the time interval from the eruption of CME1, chosen as 21 September 2009 21:00 UT, or 9.6 h simulation time after the start of shearing motions, when the centre of the flux rope crosses 1.5 R$_{\sun}$ and enters COR1-B field of view.}
        \label{fig:comparison_obs_sims}
       \end{figure*}

        In order to make the system erupt, we applied shearing motions to the southernmost polarity inversion line and modified the amplitude of the shear by adjusting the scaling factor $v_0$. Thus, a first erupting flux rope (CME1) is created approximately 8 h after the start of the shear (see Figures \ref{fig:simulations_snapshots}a,b,c), and its evolution and propagation are mostly independent of the small variations we impose on $v_0$, as seen in Figures \ref{fig:comparison_obs_sims}c,e,g. On the other hand, for the second and third flux ropes, changes within 1\% of the initial lowest shearing speed (measured at the first cell of the domain) result in three different dynamical scenarios:
        
   \begin{enumerate}
        \item The double eruption case where $|v_\phi^{max}|=\mathrm{37.4\ km\ s^{-1}}$. After the first CME, a second flux rope also erupts (CME2) from the shear applied at the boundary (see Figures \ref{fig:simulations_snapshots}d,e,f and \ref{fig:comparison_obs_sims}c,d).
        \item The stealth eruption case, where $|v_\phi^{max}|=\mathrm{37\ km\ s^{-1}}$. After the first CME, the second flux rope (FR2) created from the shearing motions falls back to the Sun (Fig. \ref{fig:simulations_snapshots}g,h,i); a third flux rope emerges from the reconfiguration of the coronal magnetic field (Fig. \ref{fig:simulations_snapshots}j,k,l) and erupts in the trail of the preceding eruption (see Fig. \ref{fig:comparison_obs_sims}e,f).
        \item The failed eruption case, where $|v_\phi^{max}|=\mathrm{37.2\ km\ s^{-1}}$. After the first CME, the second flux rope (FR2) created from the shearing motions falls back to the Sun (Fig. \ref{fig:simulations_snapshots}g,h,i); a third flux rope emerges from the reconfiguration of the coronal magnetic field (Fig. \ref{fig:simulations_snapshots}m,n,o) and reconnects with the northern closed arcade, thus not erupting (Figures \ref{fig:comparison_obs_sims}g,h).
   \end{enumerate}

Values of $|v_\phi^{max}|$ below the smallest or above the largest values of those mentioned result in only one eruption, or multiple eruptions, respectively, with the stealth occurring solely in the threshold specified. The amplitude of the applied shear is larger than the observed photospheric motions, which have typical speeds of around 5 km s$^{-1}$ \citep{manchester_shear_2007, malherbe_1983} and usually lead to flux rope formation, system destabilisation, and eventually eruption. Nevertheless, our simulated inner boundary describes the low coronal environment, and the observed magnitude of the shearing speed increases with height in the solar atmosphere, as shown by \citet {athay_1982,athay_1985} (transition region: 20 km s$^{-1}$) and by \citet{chae_shear_2000} (low corona: 20$-$50 km s$^{-1}$). Therefore, the amplitude of our imposed shear flow is in agreement with the computed velocities from coronal observations. Initially, \citet{zuccarello} and \citet{bemporad} applied the shear for 36 h in their simulations; this duration was calculated based on observed photospheric motions. In order to save computation time and resources, in this work we decrease the time interval during which the additional $v_{\phi}$ is applied from 36 h to 16 h, the latter being the value used in the discussed cases. This leads to an increase in the shear magnitude, but with almost no effect on the simulated eruptions.\\
   In all three cases listed above, the formation of the first flux rope and its eruption are essentially the same. Firstly, the applied azimuthal flow increases the magnetic pressure inside the southern arcade, making it rise and expand. Secondly, the imbalance between the magnetic pressure gradient and tension compresses the magnetic field locally. This triggers reconnection between the sides of the arcade (green circle, Fig. \ref{fig:simulations_snapshots}b) and creates the flux rope, which is deflected towards the equator owing to the southern polar magnetic pressure (Fig. \ref{fig:simulations_snapshots}a,b,c), and propagates radially inside the current sheet of the northern helmet streamer approximately after 5 R$_{\sun}$ (Fig. \ref{fig:comparison_obs_sims}). For a more detailed view of the evolution, supporting videos for all three simulations and of the observed eruptions are available in the on-line version of the article. In the simulation movies, extra flux ropes can be seen in the wake of the CMEs, and these flux ropes are somewhat comparable to those appearing in the coronagraph observations, but the occurrence of these is highly influenced by the initial coronal magnetic field configuration and the background solar wind. The magnetic configuration simulated is surely more simplified than the real configuration, and the solar wind is only adjusted such that the speed matches that measured at 1AU; therefore we do not compare these small eruptions any further since they would require simulation from a data-driven MHD model. The time in the simulation videos is not normalised to a real unit system and the conversion equation is $t_{real}=t_{sim} \times 1.9325$ h, where $t_{sim}$ is the time shown in movies. Also, the distances on axes are expressed in solar radii. The movie of the observations was visualised in the JHelioviewer software, using running difference images from the EUVI, COR1, and COR2 instruments on board STEREO-B. \\ 
      
   The fastest shear we applied ($\mathrm{37.4\ km\ s^{-1}}$) results in the formation of another flux rope in the wake of the first CME, much in the same manner as the previous flux rope. The arcade expands again, and the sides pinch (green circle, Fig. \ref{fig:simulations_snapshots}d) and create the flux rope, which then erupts into the equatorial current sheet to produce CME2 (Fig. \ref{fig:simulations_snapshots}d,e,f). The evolution and propagation can be visualised in Fig. \ref{fig:comparison_obs_sims}c,d. This second flux rope also forms in the next two scenarios, but it fails to erupt and falls back to the Sun, dissipating at the boundary (Fig. \ref{fig:simulations_snapshots}g,h,i). \\
   The lowest speed we used ($\mathrm{37\ km\ s^{-1}}$) produced CME1 in the same way, but the magnetic flux introduced was insufficient to inflate the arcade  long enough to make it detach once more. Instead, the unstable current sheet formed in the trail of CME1 reconnects (green circle, Fig. \ref{fig:simulations_snapshots}j) and creates a third flux rope after $\approx$17.5 h of shearing, which is then slowly deflected towards equator (Fig. \ref{fig:simulations_snapshots}j,k,l). Given the high altitude at which the reconnection sets in (1.4 R$_{\sun}$) and the low speed of the process, this second eruption most likely does not leave any clear signatures, placing it in the stealth CME category \citep{robbrecht_stealth}. Compared to CME2 from the previous case, this one is narrower, slower, and occurs later, as seen in Figures \ref{fig:comparison_obs_sims}e,f. \\
   Applying an intermediate shearing speed ($\mathrm{37.2\ km\ s^{-1}}$) again gives rise to CME1, to a second flux rope that falls back to the Sun, and to a third flux rope emerging through the same mechanism as the stealth, from coronal magnetic field reconnection (green circle, Fig. \ref{fig:simulations_snapshots}m) $\approx$19 h after the start of the shearing. Its evolution, however, is different since the small amount of additional flux translates into a slightly more energetic reconfiguration, which increases the speed of the flux rope. As a consequence of its higher acceleration, slightly higher speed, and lower formation height (1.55 R$_{\sun}$, taken at the centre of the flux rope, as opposed to 1.7 R$_{\sun}$ for the stealth), the CME cannot be deflected as easily towards the equator by the magnetic pressure gradient, and eventually reconnects with the large northern arcade, creating a failed (confined) eruption (Fig. \ref{fig:simulations_snapshots}m,n,o). We analysed and compared the dynamics of the three scenarios in more detail, and these results are presented in the next section.


        \section{Analysis of the simulations}
        
\begin{figure}[h!]
                        \centering
                        \resizebox{\hsize}{!}{\includegraphics{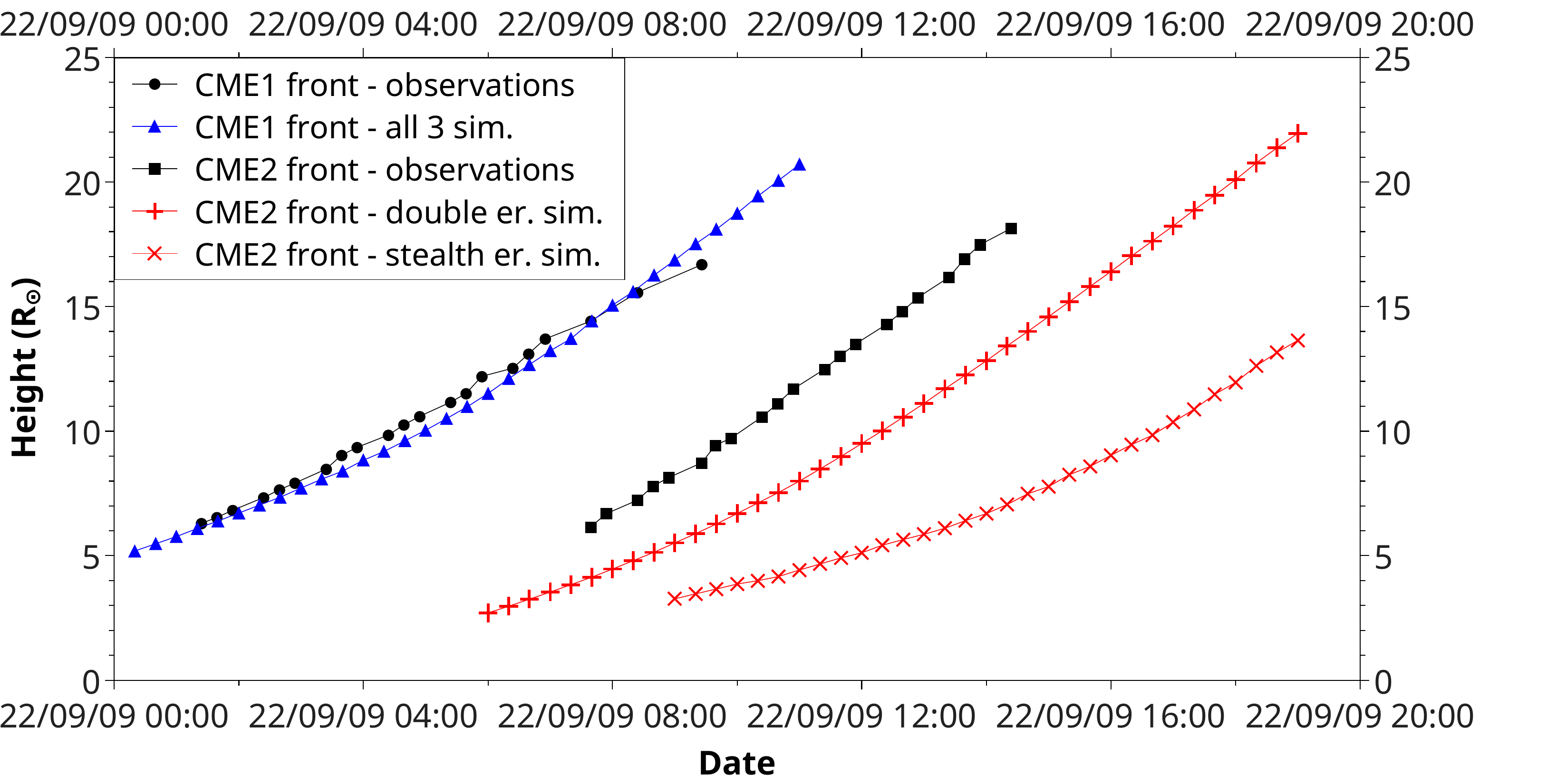}}
                        \caption{Location of the CMEs fronts in the meridional plane as a function of time. The black dots indicate the observed CME1, black squares indicate the observed CME2, blue triangles represent the simulated CME1, red crosses indicate the simulated CME2 (double eruption scenario), and red diagonal crosses represent the simulated stealthy CME2.}
                        \label{fig:ht_plots_sims_obs}   
                \end{figure}

        In order to compare the dynamics of the simulated and observed CMEs, we plotted in the deprojected heights of the eruption fronts and the locations of simulated leading edges in Fig. \ref{fig:ht_plots_sims_obs}. The real distances from Sun to the observed CMEs were required in this step due to the 2.5D topology of the simulated eruptions, which would exclude any longitudinal deflection. The CMEs on 21-22 September 2009 were tracked only in the COR2-B field of view, from which they began to propagate radially. Below 6 R$_{\sun}$, the strong deflection and diffuse fronts of the CMEs prevented us from extracting accurate information.\\
        In the simulation data shown in Figure \ref{fig:ht_plots_sims_obs} there is only one representation of the first CME because it evolved similarly for all three cases, as described in Section \ref{sec:results}. We note that the first simulated eruption was slightly faster than the observed eruption, with a speed of $\mathrm{279\ km\ s^{-1}}$ corresponding to the last measured height in COR2-B, where the speed was $\mathrm{257\ km\ s^{-1}}$. Given the fact that the background magnetic configuration and solar wind were not perfectly simulated to match remote sensing and in situ observations, this value is a good approximation. The second CME of the double eruption case correlates better with its observed counterpart, which has a simulated speed of $\mathrm{342\ km\ s^{-1}}$ at the last data point at a height of 18 R$_{\sun}$, as compared to the observed $\mathrm{349\ km\ s^{-1}}$. The stealth CME front was starting to merge with the trail of the first eruption at that distance, and therefore we were unable to detect it, but the speed of the centre of the flux rope was $\mathrm{339\ km\ s^{-1}}$. All three values closely resemble the observed values, therefore managing to realistically reproduce two slow CMEs, with two different erupting scenarios.               
        
                \begin{figure}[h!]
                        \centering
                        \resizebox{\hsize}{!}{\includegraphics{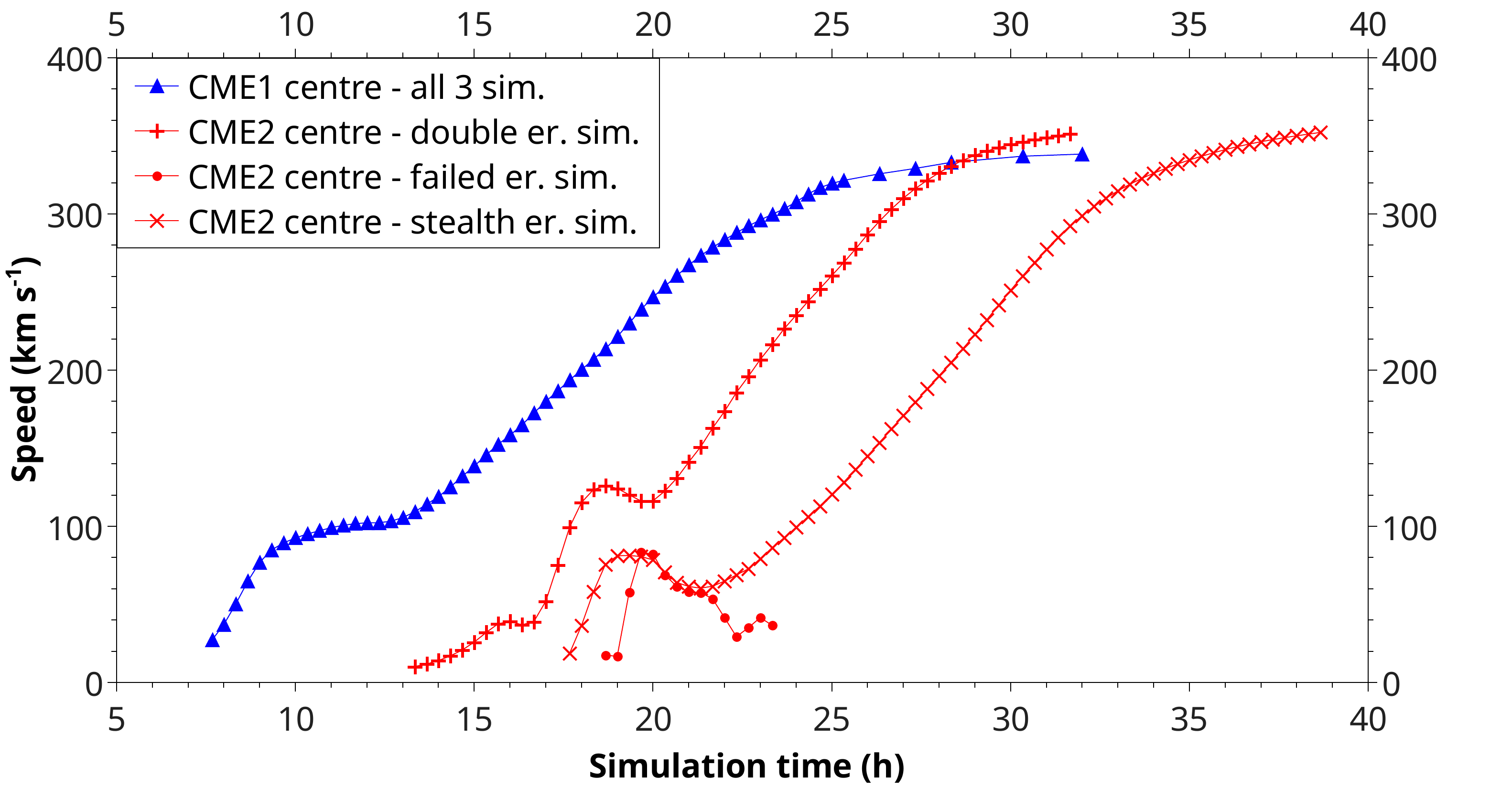}}
                        \caption{Total speed of the simulated CMEs, calculated at the centre of the flux rope. The blue triangles indicate CME1, the red crosses indicate CME2 (double eruption scenario), the red dots represent CME2 (failed eruption), and the red diagonal crosses indicate stealthy CME2.}
                        \label{fig:speed-sims}  
                \end{figure}
        
                \begin{figure}[h!]
                        \centering
                        \resizebox{\hsize}{!}{\includegraphics{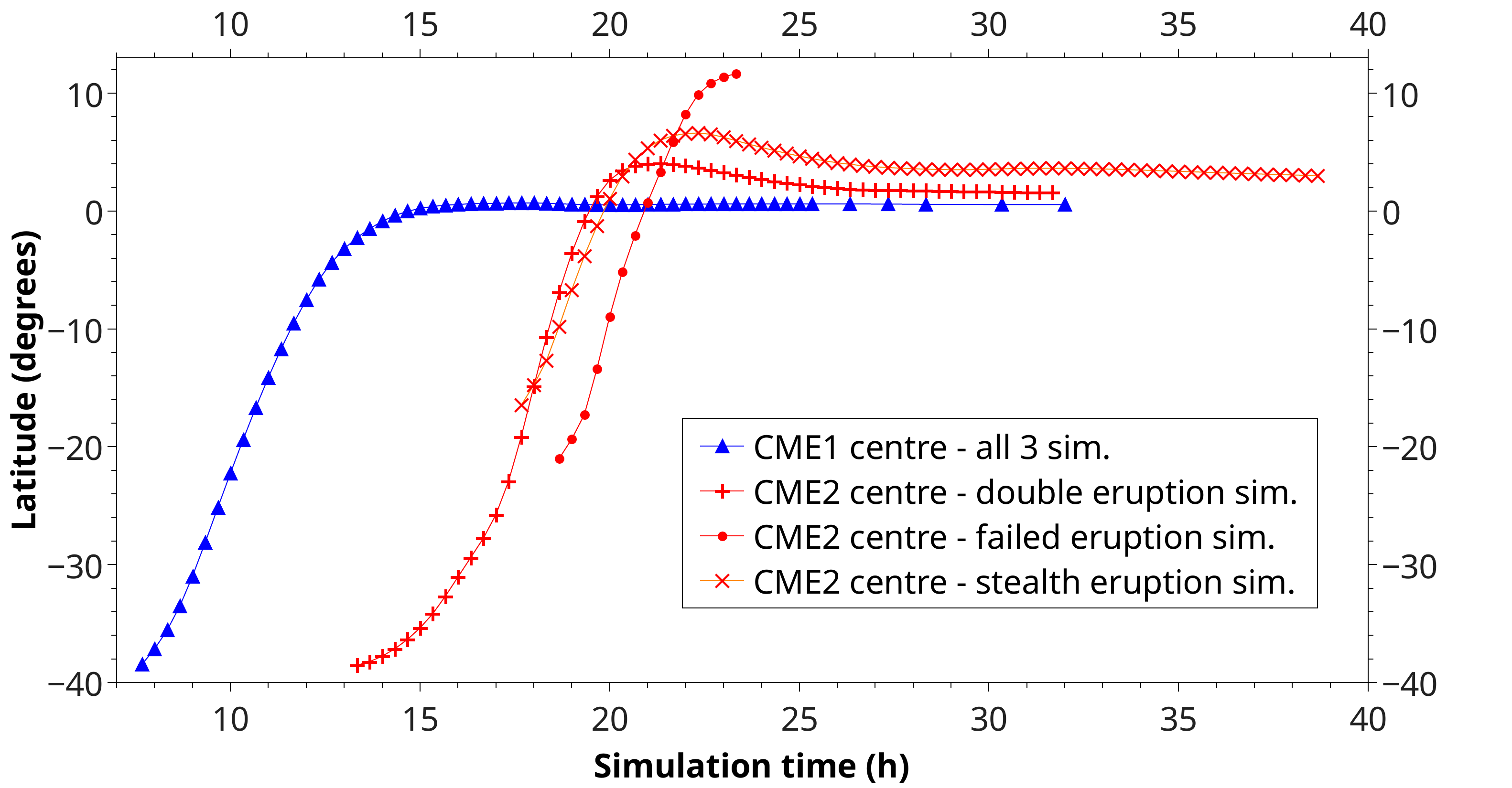}}
                        \caption{Latitude of the simulated CMEs with respect to the Sun's equator as a function of time, taken at the centre of the flux rope. The legend is the same as in Figure \ref{fig:speed-sims}.}
                        \label{fig:lat-sims}    
                \end{figure}

\begin{figure}[h!]
                        \centering
                        \resizebox{\hsize}{!}{\includegraphics{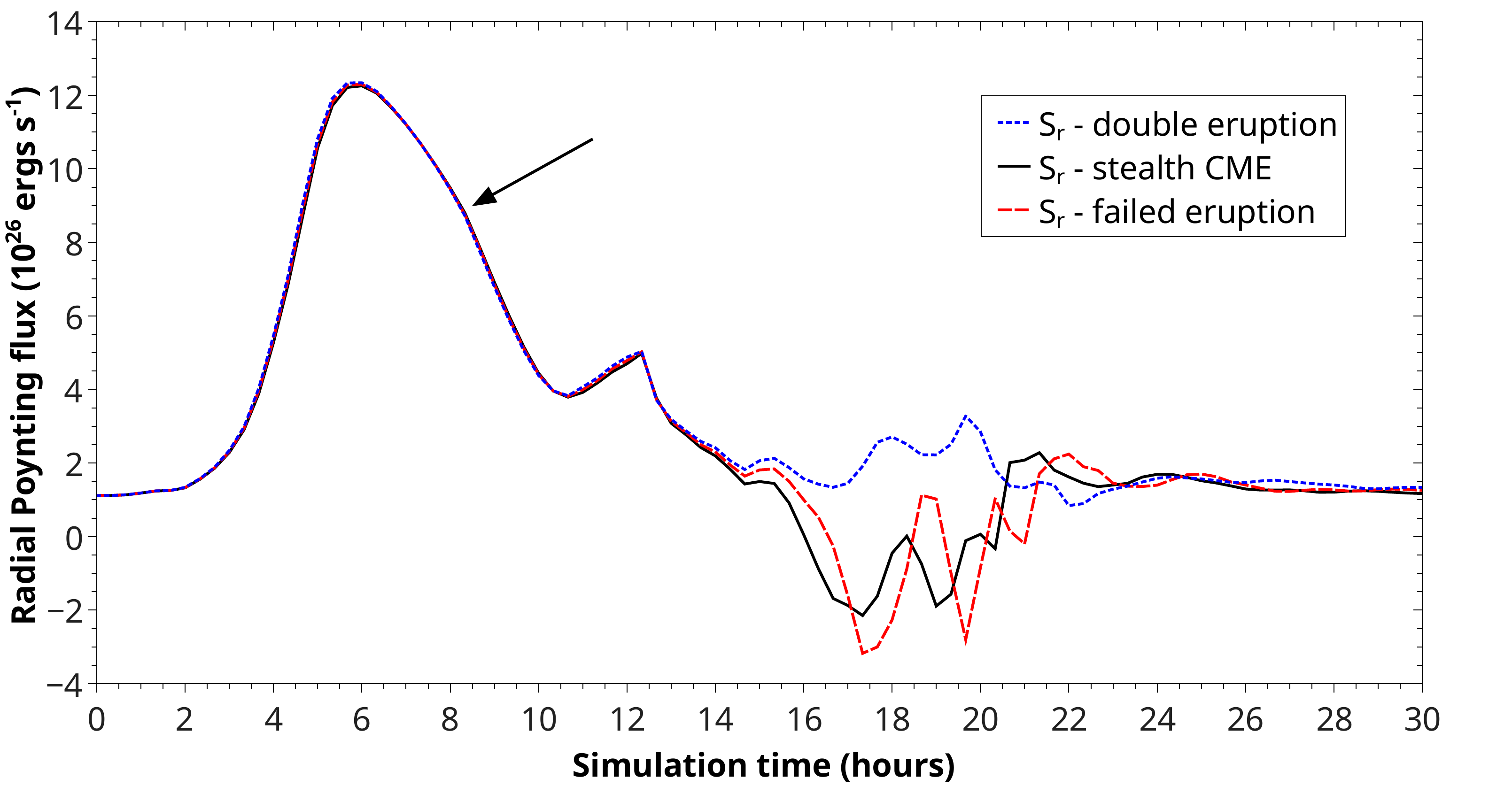}}
                        \caption{Evolution in time of the radial component of the Poynting flux, summed over the whole Sun.}
                        \label{fig:s_r}
                \end{figure}

    Next, we focussed more on the simulations as an independent result, rather than comparing them to the observations, and we analysed the differences between the three cases. As shown in Figure \ref{fig:speed-sims}, all three secondary eruptions present a stronger acceleration compared to CME1 when close to the Sun, owing to the depletion of background wind, and to the already open field lines created from the passage of the first CME. We note that CME2 from the double eruption case presents a slow initial rising phase. All three of these secondary eruptions emerge in the current sheet created by the first eruption, whereas CME1 needs to overcome the magnetic tension of the overlying field, hence the smaller acceleration and slow rise. The higher initial acceleration makes it harder for these narrower flux ropes to be deflected towards the equator, so they reach higher latitudes than CME1, as shown in Figure \ref{fig:lat-sims}. In the case of the failed eruption, the slightly larger momentum of the flux rope increases the latitudinal deflection, making it unable to escape the reconnection with the northern arcade. The timeline of its evolution can be seen in Fig. \ref{fig:simulations_snapshots}m,n,o, and is described as follows: 20.5 h after the initiation of the shearing motions, the flux rope enters the northern streamer; 20 min later, it gets trapped by the closed arcades, and after 23.5 h, it reconnects entirely and disappears. In the double eruption scenario, the second flux rope begins to form around 13.5 h after the start of the shearing motions, and is accelerated away from the Sun 3 h later. In the case of the stealth CME, the flux rope becomes embedded into the streamer after 19 h (Fig. \ref{fig:simulations_snapshots}k), a step which induces deceleration for all the scenarios, at their corresponding times. \\  
        Upon reaching the highest latitude, the polar magnetic pressure together with the northern streamer force all CMEs to deflect equatorward, thus also contributing to the deceleration of the CMEs. Once the propagation becomes radial and the CMEs enter the solar wind, they accelerate again and reach speeds of up to 350km/s. Even though the CME formation and cause of eruption are different, it is interesting to point out the similarities in the overall speed profile of CME2 for the stealth and double eruption scenarios.\\
        We also performed Poynting flux analysis by calculating the radial component of its associated vector close to the inner boundary through a sphere located at 1.14 R$_{\sun}$ as follows:
        
        \begin{equation}
        S_r =\frac{[-(\boldsymbol{v} \times \boldsymbol{B}) \times \boldsymbol{B}]_r}{\mu_0} =\frac{B^2 v_r-(\boldsymbol{v} \cdot \boldsymbol{B})B_r}{\mu_0},  
        \end{equation}

        \noindent and multiplying every value with the corresponding area of the surface obtained by rotating the bottom of the grid cell around the Sun at 1.14 R$_{\sun}$. The final values consist in the summation over the entire surface of the shell around the Sun, and their variation in time is shown in Figure \ref{fig:s_r}, where time=0 indicates the start of the shear. As expected, the induced shearing motions increase the electromagnetic energy flow through the inner boundary, as seen in the first 6 h of simulation, during which the southern arcade starts expanding. Afterwards, the lateral magnetic pressure gradient compresses the field locally to create the first flux rope, and after $\approx$ 8 h this flux rope detaches from the Sun as it is accelerated a first time, which corresponds to the `shoulder' in the radial Poynting flux, indicated by the arrow in Fig. \ref{fig:s_r}. Following the eruption of CME1, a second flux rope begins to form as a consequence of ongoing shearing motions and lateral pinching commences after the peak at 12 h. This evolution is the same for all three simulation cases within the first 14 h. In the next part, the dynamics start to differentiate because this recently created flux rope erupts and forms CME2 in the double eruption case. For the other two scenarios, the remaining shear does not impose sufficient energy onto the flux rope to overcome the magnetic tension of the overlying closed magnetic field, so the flux rope falls back to the Sun at $\approx$ 15 h and disperses at the inner boundary, as described in Section \ref{sec:results}.\\
        A particularly interesting aspect presented in Figure \ref{fig:s_r} is that the Poynting flux profile is practically the same until near the end of the shearing motions (which were applied for 16 h), yet the coronal environment behaves completely different from that point onwards. Furthermore, the radial electromagnetic energy flux is very similar in the cases of stealth and failed eruptions, emphasising the argument stated before that these flux ropes are not caused by the shear, instead they result from the coronal magnetic field reconfiguration. This can also be seen in Figure \ref{fig:speed-sims}, in the difference between initiation phases, which presents a slow rise for CME2 in the double eruption case, whereas  for the other two, which form higher in the corona, the moment they are created also coincides to the initial acceleration stage.                     
                
        \section{Summary}
        
        In this paper, the initial observational basis consisted in a multiple CME event on 21-22 September 2009, with an interval between the eruptions of approximately eight hours. In order to obtain the real propagation directions, we performed GCS analysis high in the corona on both CMEs, and derived an average longitude of 5.82$\degr$W for CME1 and of 6.7$\degr$W for CME2, respectively.\\   
        Subsequently, we used the MHD package of the code MPI-AMRVAC to numerically simulate consecutive CMEs by applying shearing motions at the inner boundary, starting from an initial triple arcade structure embedded into a bimodal solar wind. We changed the amplitude of the shear to understand the effects of its variation on the induced eruptions with the goal of obtaining a stealth eruption in the trailing current sheet of a preceding CME. These new results indicate that within just 1\% variation of the lowest shearing speed (37 $\mathrm{km\ s^{-1}}$), three different eruption scenarios for the second CME were created: a stealth, a failed, and a double eruption. We emphasise in particular the initiation, for the first time, of a stealth from the change in applied shear, and not only from the overlying magnetic field constraints, as previously studied. All three cases are comprised of a first eruption generated from the imposed stress at the inner boundary, and a second eruption occurring either from the reconfiguration of coronal magnetic field or from the additional $v_\phi$ component. We compared the height-time evolution of these eruptions with the observations described above, and achieved a good slope correlation for CME1 and CME2 from the double eruption scenario. The simulated deflection of both ejections in all cases was also studied by tracking the latitude and total speed of the flux rope centres. We would like to emphasise the physical interpretation of these results, in the sense that the solar corona can react and produce eruptions differently, given almost the same initial triggering factors, making the predictions of such eruptive processes even more difficult.\\
        We also investigated the evolution of the radial Poynting flux close to the inner boundary, to acquire a global perspective of the electromagnetic energy introduced into the domain. The steps in dynamical evolution of the simulated CMEs could also be correlated with certain features, such as maximum, minimum or fluctuations, in the Poynting flux profile. This highlights the surprising similarities between the stealth and failed eruptions and the sensitive coronal response to stress factors. In a follow-up paper we will discuss the propagation of these simulations in the interplanetary space and compare their signatures to in situ measurements.

        
        \begin{acknowledgements}
             We thank the referee for their constructive comments and suggestions, which led to the improvement of this article. 
                 D.C.T. was funded by the Ph.D. fellowship of the Research Foundation – Flanders (FWO), contract number T1 1118918N. D.C.T., E.D. and M.M. acknowledge support from the Belgian Federal Science Policy Office (BELSPO) in the framework of the ESA-PRODEX program, grant No. 4000120800. This work was partially supported by a PhD Grant awarded by the Royal Observatory of Belgium.  E.C. was funded by the Research Foundation - Flanders (grant FWO 12M0119N).These results were also obtained in the framework of the projects C14/19/089 (C1 project Internal Funds KU Leuven), G.0A23.16N (FWO-Vlaanderen), and C 90347 (ESA Prodex). For the computations we used the infrastructure of the VSC – Flemish Supercomputer Center, funded by the Hercules foundation and the Flemish Government – department EWI.
                 We acknowledge use of NASA/GSFC's Space Physics Data Facility's OMNIWeb (or CDAWeb or ftp) service, and OMNI data. We acknowledge the usage of JHelioviewer software \citep{jhelioviewer} and the use of STEREO/SECCHI data (http://stereo-ssc.nascom.nasa.gov/).

        \end{acknowledgements}

        \bibliographystyle{aa}
        \bibliography{talpeanu.bib}

\begin{thebibliography}{48}
\expandafter\ifx\csname natexlab\endcsname\relax\def\natexlab#1{#1}\fi

\bibitem[{Alzate \& Morgan(2017)}]{alzate_sources}
Alzate, N. \& Morgan, H. 2017, The Astrophysical Journal, 840, 103

\bibitem[{{Athay} {et~al.}(1982){Athay}, {Gurman}, {Shine}, \&
  {Henze}}]{athay_1982}
{Athay}, R.~G., {Gurman}, J.~B., {Shine}, R.~A., \& {Henze}, W. 1982, \apj,
  261, 684

\bibitem[{{Athay} {et~al.}(1985){Athay}, {Jones}, \& {Zirin}}]{athay_1985}
{Athay}, R.~G., {Jones}, H.~P., \& {Zirin}, H. 1985, \apj, 288, 363

\bibitem[{{Bemporad} {et~al.}(2012){Bemporad}, {Zuccarello}, {Jacobs},
  {Mierla}, \& {Poedts}}]{bemporad}
{Bemporad}, A., {Zuccarello}, F.~P., {Jacobs}, C., {Mierla}, M., \& {Poedts},
  S. 2012, \solphys, 281, 223

\bibitem[{Chae {et~al.}(2000)Chae, Wang, Qiu, Goode, \&
  Wilhelm}]{chae_shear_2000}
Chae, J., Wang, H., Qiu, J., Goode, P.~R., \& Wilhelm, K. 2000, The
  Astrophysical Journal, 533, 535

\bibitem[{{Chan{\'e}} {et~al.}(2008){Chan{\'e}}, {Poedts}, \& {van der
  Holst}}]{chane_2008}
{Chan{\'e}}, E., {Poedts}, S., \& {van der Holst}, B. 2008, \aap, 492, L29

\bibitem[{{Chan{\'e}} {et~al.}(2006){Chan{\'e}}, {van der Holst}, {Jacobs},
  {Poedts}, \& {Kimpe}}]{chane_2006}
{Chan{\'e}}, E., {van der Holst}, B., {Jacobs}, C., {Poedts}, S., \& {Kimpe},
  D. 2006, \aap, 447, 727

\bibitem[{Dedner {et~al.}(2002)Dedner, Kemm, Kröner, Munz, Schnitzer, \&
  Wesenberg}]{glm}
Dedner, A., Kemm, F., Kröner, D., {et~al.} 2002, Journal of Computational
  Physics, 175, 645

\bibitem[{DeVore \& Antiochos(2000)}]{devore_shear}
DeVore, C.~R. \& Antiochos, S.~K. 2000, The Astrophysical Journal, 539, 954

\bibitem[{{D'Huys} {et~al.}(2014){D'Huys}, {Seaton}, {Poedts}, \&
  {Berghmans}}]{elke_stealth}
{D'Huys}, E., {Seaton}, D.~B., {Poedts}, S., \& {Berghmans}, D. 2014, \apj,
  795, 49

\bibitem[{Forbes(2000)}]{forbes_energies}
Forbes, T.~G. 2000, Journal of Geophysical Research: Space Physics, 105, 23153

\bibitem[{{Groth} {et~al.}(2000){Groth}, {De Zeeuw}, {Gombosi}, \&
  {Powell}}]{groth}
{Groth}, C.~P.~T., {De Zeeuw}, D.~L., {Gombosi}, T.~I., \& {Powell}, K.~G.
  2000, \jgr, 105, 25053

\bibitem[{{Hosteaux} {et~al.}(2018){Hosteaux}, {Chan{\'e}}, {Decraemer},
  {Talpeanu}, \& {Poedts}}]{skralan}
{Hosteaux}, S., {Chan{\'e}}, E., {Decraemer}, B., {Talpeanu}, D.~C., \&
  {Poedts}, S. 2018, \aap, 620, A57

\bibitem[{{Hosteaux} {et~al.}(2019){Hosteaux}, {Chan{\'e}}, \&
  {Poedts}}]{skralan_2019}
{Hosteaux}, S., {Chan{\'e}}, E., \& {Poedts}, S. 2019, arXiv e-prints,
  arXiv:1910.04680

\bibitem[{{Howard} {et~al.}(2008){Howard}, {Moses}, {Vourlidas}, {Newmark},
  {Socker}, {Plunkett}, {Korendyke}, {Cook}, {Hurley}, {Davila}, {Thompson},
  {St Cyr}, {Mentzell}, {Mehalick}, {Lemen}, {Wuelser}, {Duncan}, {Tarbell},
  {Wolfson}, {Moore}, {Harrison}, {Waltham}, {Lang}, {Davis}, {Eyles},
  {Mapson-Menard}, {Simnett}, {Halain}, {Defise}, {Mazy}, {Rochus}, {Mercier},
  {Ravet}, {Delmotte}, {Auchere}, {Delaboudiniere}, {Bothmer}, {Deutsch},
  {Wang}, {Rich}, {Cooper}, {Stephens}, {Maahs}, {Baugh}, {McMullin}, \&
  {Carter}}]{stereo_secchi}
{Howard}, R.~A., {Moses}, J.~D., {Vourlidas}, A., {et~al.} 2008, \ssr, 136, 67

\bibitem[{{Howard} \& {Harrison}(2013)}]{howard_harrison_review}
{Howard}, T.~A. \& {Harrison}, R.~A. 2013, \solphys, 285, 269

\bibitem[{Jacobs {et~al.}(2005)Jacobs, Poedts, van~der Holst, \&
  Chané}]{jacobs_2005}
Jacobs, C., Poedts, S., van~der Holst, B., \& Chané, E. 2005,
  http://dx.doi.org/10.1051/0004-6361:20041676, 430

\bibitem[{{Kaiser} {et~al.}(2008){Kaiser}, {Kucera}, {Davila}, {St.~Cyr},
  {Guhathakurta}, \& {Christian}}]{stereo_mission}
{Kaiser}, M.~L., {Kucera}, T.~A., {Davila}, J.~M., {et~al.} 2008, \ssr, 136, 5

\bibitem[{{Karpen} {et~al.}(2012){Karpen}, {Antiochos}, \& {DeVore}}]{karpen}
{Karpen}, J.~T., {Antiochos}, S.~K., \& {DeVore}, C.~R. 2012, \apj, 760, 81

\bibitem[{Keppens {et~al.}(2012)Keppens, Meliani, van Marle, Delmont, Vlasis,
  \& van~der Holst}]{keppens_amrvac}
Keppens, R., Meliani, Z., van Marle, A., {et~al.} 2012, Journal of
  Computational Physics, 231, 718 , special Issue: Computational Plasma Physics

\bibitem[{{Kilpua} {et~al.}(2014){Kilpua}, {Mierla}, {Zhukov}, {Rodriguez},
  {Vourlidas}, \& {Wood}}]{kilpua_stealth}
{Kilpua}, E.~K.~J., {Mierla}, M., {Zhukov}, A.~N., {et~al.} 2014, \solphys,
  289, 3773

\bibitem[{{Ko} {et~al.}(2003){Ko}, {Raymond}, {Lin}, {Lawrence}, {Li}, \&
  {Fludra}}]{ko_blobs}
{Ko}, Y.-K., {Raymond}, J.~C., {Lin}, J., {et~al.} 2003, \apj, 594, 1068

\bibitem[{{Linker} \& {Mikic}(1995)}]{linker_shear}
{Linker}, J.~A. \& {Mikic}, Z. 1995, \apjl, 438, L45

\bibitem[{Lynch {et~al.}(2016)Lynch, Masson, Li, DeVore, Luhmann, Antiochos, \&
  Fisher}]{lynch_stealth}
Lynch, B.~J., Masson, S., Li, Y., {et~al.} 2016, Journal of Geophysical
  Research: Space Physics, 121, 10,677

\bibitem[{Ma {et~al.}(2010)Ma, Attrill, Golub, \& Lin}]{ma_stealth}
Ma, S., Attrill, G. D.~R., Golub, L., \& Lin, J. 2010, The Astrophysical
  Journal, 722, 289

\bibitem[{{Malherbe} {et~al.}(1983){Malherbe}, {Schmieder}, {Ribes}, \&
  {Mein}}]{malherbe_1983}
{Malherbe}, J.~M., {Schmieder}, B., {Ribes}, E., \& {Mein}, P. 1983, \aap, 119,
  197

\bibitem[{Manchester(2007)}]{manchester_shear_2007}
Manchester, W.~I. 2007, The Astrophysical Journal, 666, 532

\bibitem[{Manchester~IV {et~al.}(2004)Manchester~IV, Gombosi, Roussev, Ridley,
  De~Zeeuw, Sokolov, Powell, \& Tóth}]{manchester}
Manchester~IV, W.~B., Gombosi, T.~I., Roussev, I., {et~al.} 2004, Journal of
  Geophysical Research: Space Physics, 109
  [\eprint{https://agupubs.onlinelibrary.wiley.com/doi/pdf/10.1029/2003JA010150}]

\bibitem[{{M\"uller, D.} {et~al.}(2017){M\"uller, D.}, {Nicula, B.}, {Felix,
  S.}, {Verstringe, F.}, {Bourgoignie, B.}, {Csillaghy, A.}, {Berghmans, D.},
  {Jiggens, P.}, {Garc\'{\i}a-Ortiz, J. P.}, {Ireland, J.}, {Zahniy, S.}, \&
  {Fleck, B.}}]{jhelioviewer}
{M\"uller, D.}, {Nicula, B.}, {Felix, S.}, {et~al.} 2017, A\&A, 606, A10

\bibitem[{Nitta \& Mulligan(2017)}]{nariaki_stealth}
Nitta, N.~V. \& Mulligan, T. 2017, Solar Physics, 292, 125

\bibitem[{Pevtsov {et~al.}(2012)Pevtsov, Panasenco, \&
  Martin}]{pevtsov_filament}
Pevtsov, A.~A., Panasenco, O., \& Martin, S.~F. 2012, Solar Physics, 277, 185

\bibitem[{Porth {et~al.}(2014)Porth, Xia, Hendrix, Moschou, \&
  Keppens}]{porth_amrvac}
Porth, O., Xia, C., Hendrix, T., Moschou, S.~P., \& Keppens, R. 2014, The
  Astrophysical Journal Supplement Series, 214, 4

\bibitem[{{Riley} {et~al.}(2007){Riley}, {Lionello}, {Miki{\'c}}, {Linker},
  {Clark}, {Lin}, \& {Ko}}]{riley_sim_blobs}
{Riley}, P., {Lionello}, R., {Miki{\'c}}, Z., {et~al.} 2007, \apj, 655, 591

\bibitem[{{Robbrecht} {et~al.}(2009{\natexlab{a}}){Robbrecht}, {Berghmans}, \&
  {Van der Linden}}]{cactus}
{Robbrecht}, E., {Berghmans}, D., \& {Van der Linden}, R.~A.~M.
  2009{\natexlab{a}}, \apj, 691, 1222

\bibitem[{{Robbrecht} {et~al.}(2009{\natexlab{b}}){Robbrecht}, {Patsourakos},
  \& {Vourlidas}}]{robbrecht_stealth}
{Robbrecht}, E., {Patsourakos}, S., \& {Vourlidas}, A. 2009{\natexlab{b}},
  \apj, 701, 283

\bibitem[{Thernisien(2011)}]{gcs_thernisien_2011}
Thernisien, A. 2011, The Astrophysical Journal Supplement Series, 194, 33

\bibitem[{{Thernisien} {et~al.}(2009){Thernisien}, {Vourlidas}, \&
  {Howard}}]{gcs_thernisien_2009}
{Thernisien}, A., {Vourlidas}, A., \& {Howard}, R.~A. 2009, \solphys, 256, 111

\bibitem[{{Thernisien} {et~al.}(2006){Thernisien}, {Howard}, \&
  {Vourlidas}}]{gcs_thernisien_2006}
{Thernisien}, A.~F.~R., {Howard}, R.~A., \& {Vourlidas}, A. 2006, \apj, 652,
  763

\bibitem[{Thompson {et~al.}(2003)Thompson, Davila, R.~Fisher, E.~Orwig,
  E.~Mentzell, E.~Hetherington, J.~Derro, E.~Federline, C.~Clark, T.~C.~Chen,
  L.~Tveekrem, J.~Martino, Novello, Wesenberg, C.~StCyr, Reginald, Howard,
  I.~Mehalick, J.~Hersh, \& Elmore}]{thompson-cor}
Thompson, W., Davila, J., R.~Fisher, R., {et~al.} 2003, Proc SPIE, 1

\bibitem[{{Thompson}(2006)}]{thompson_stonyhurst}
{Thompson}, W.~T. 2006, \aap, 449, 791

\bibitem[{Van~der Holst {et~al.}(2007)Van~der Holst, Jacobs, \&
  Poedts}]{bart_breakout}
Van~der Holst, B., Jacobs, C., \& Poedts, S. 2007, The Astrophysical Journal,
  671, L77

\bibitem[{Van~der Holst {et~al.}(2006)Van~der Holst, Poedts, Chan{\'e}, Jacobs,
  Dubey, \& Kimpe}]{bart_shear}
Van~der Holst, B., Poedts, S., Chan{\'e}, E., {et~al.} 2006, Space Science
  Reviews, 121, 91

\bibitem[{{Webb} \& {Cliver}(1995)}]{webb_and_cliver_blobs}
{Webb}, D.~F. \& {Cliver}, E.~W. 1995, \jgr, 100, 5853

\bibitem[{Webb \& Howard(2012)}]{webb_speeds}
Webb, D.~F. \& Howard, T.~A. 2012, Living Reviews in Solar Physics, 9, 3

\bibitem[{Webb \& Vourlidas(2016)}]{webb_and_vourlidas_blobs}
Webb, D.~F. \& Vourlidas, A. 2016, Solar Physics, 291, 3725

\bibitem[{{Wuelser} {et~al.}(2004){Wuelser}, {Lemen}, {Tarbell}, {Wolfson},
  {Cannon}, {Carpenter}, {Duncan}, {Gradwohl}, {Meyer}, {Moore}, {Navarro},
  {Pearson}, {Rossi}, {Springer}, {Howard}, {Moses}, {Newmark},
  {Delaboudiniere}, {Artzner}, {Auchere}, {Bougnet}, {Bouyries}, {Bridou},
  {Clotaire}, {Colas}, {Delmotte}, {Jerome}, {Lamare}, {Mercier}, {Mullot},
  {Ravet}, {Song}, {Bothmer}, \& {Deutsch}}]{secchi}
{Wuelser}, J.-P., {Lemen}, J.~R., {Tarbell}, T.~D., {et~al.} 2004, in
  \procspie, Vol. 5171, Telescopes and Instrumentation for Solar Astrophysics,
  ed. S.~{Fineschi} \& M.~A. {Gummin}, 111--122

\bibitem[{Xia {et~al.}(2018)Xia, Teunissen, Mellah, Chan{\'{e}}, \&
  Keppens}]{xia_amrvac}
Xia, C., Teunissen, J., Mellah, I.~E., Chan{\'{e}}, E., \& Keppens, R. 2018,
  The Astrophysical Journal Supplement Series, 234, 30

\bibitem[{{Zuccarello} {et~al.}(2012){Zuccarello}, {Bemporad}, {Jacobs},
  {Mierla}, {Poedts}, \& {Zuccarello}}]{zuccarello}
{Zuccarello}, F.~P., {Bemporad}, A., {Jacobs}, C., {et~al.} 2012, \apj, 744, 66

\end{thebibliography}
        
\end{document}